\documentclass[12pt,a4paper]{article}
\pdfoutput=1

\usepackage{amsmath}
\usepackage{amssymb}
\usepackage{mathrsfs}
\usepackage{graphicx}
\usepackage{enumerate}

\usepackage[margin=1in]{geometry}
\usepackage[pdftex,colorlinks=true,linkcolor=blue,citecolor=blue,urlcolor=blue]{hyperref}

\graphicspath{{Figures/}}

\newcommand{\eqpic}[2]{\vcenter{\hbox{\includegraphics[height=#2ex]{#1}}}}

\newcommand{\z}{\mathcal{Z}}
\newcommand{\bulk}{\mathcal{M}}
\newcommand{\rs}{X} 
\newcommand{\tzero}{\Sigma} 

\newcommand{\ZZ}{\mathbb{Z}}
\newcommand{\RR}{\mathbb{R}}
\newcommand{\CC}{\mathbb{C}}
\newcommand{\PP}{\mathbb{P}}
\newcommand{\HH}{\mathbb{H}}
\newcommand{\dd}{\mathrm{d}}

\DeclareMathOperator{\csch}{csch}
\DeclareMathOperator{\tr}{Tr}

\begin{document}

\title{Holographic partition functions and phases for higher genus Riemann surfaces}
\author{Henry Maxfield$^{1}$\footnote{henry.maxfield@physics.mcgill.ca},{ } Simon F. Ross$^{2}$\footnote{s.f.ross@durham.ac.uk}{ } and Benson Way$^{3}$ \footnote{bw356@cam.ac.uk}\\  \bigskip \\
$^{1}$McGill Physics Department, 3600 rue University, \\ Montr\'eal, QC H3A 2T8, Canada
\\
\\
$^{2}$Centre for Particle Theory, Department of Mathematical Sciences \\ Durham University\\ South Road, Durham DH1 3LE
\\
\\
$^{3}$DAMTP, Centre for Mathematical Sciences, University of Cambridge \\ Wilberforce Road, Cambridge }

\maketitle

\vspace{-1em}

\abstract{
We describe a numerical method to compute the action of Euclidean saddle-points for the partition function of a two-dimensional holographic CFT on a Riemann surface of arbitrary genus, with constant curvature metric. We explicitly evaluate the action for the saddles for genus two and map out the phase structure of dominant bulk saddles in a two-dimensional subspace of the moduli space. We discuss spontaneous breaking of discrete symmetries, and show that the handlebody bulk saddles always dominate over certain non-handlebody solutions.
}

\clearpage

\tableofcontents

\section{Introduction}

The relationship between two-dimensional conformal field theories and gravity in AdS$_3$ is one of the most interesting arenas for exploring and testing the AdS/CFT correspondence, both in real Lorentzian spacetimes and for the Euclidean partition function. The presence of an infinite-dimensional conformal symmetry on the field theory and the simplicity of gravity in three dimensions make it possible to carry out calculations that are inaccessible in higher dimensions. Due to the equivalence of conformal structure in two dimensions and complex structure, it is natural to study a CFT on a Riemann surface, and the simplest quantity to compute is the path integral on the surface, which defines the partition function. For the torus and higher genus surfaces, there is a nontrivial (though finite dimensional) moduli space, and one interesting avenue of investigation is the study of the dependence of the partition function on the moduli. For a holographic CFT in the semiclassical limit, this is computed by the action of a geometry solving the equations of motion, which for pure gravity is a hyperbolic three-manifold with the given surface as its asymptotic boundary, due to the local triviality of Einstein's equations in three dimensions. For a given Riemann surface on the boundary, there are many possible on-shell geometries which can contribute as saddle points to the partition function, including handlebody solutions, as well as non-handlebodies. As the moduli vary, there are phase transitions in the partition function where bulk saddle-points exchange dominance, generalising the Hawking-Page transition.

For the torus, this problem has been extensively explored, and the full partition function can be related to a sum over geometries, related by the modular group \cite{Dijkgraaf:2000fq,Manschot:2007ha,Witten:2007kt,Maloney:2007ud,Keller:2014xba}. The situation for higher genus Riemann surfaces has been less thoroughly explored, although the first steps were taken more than a decade ago \cite{Krasnov:2000zq,Krasnov:2003ye}. The main barrier to progress in the higher genus case was that although \cite{Krasnov:2000zq} reduced the evaluation of the action for a bulk handlebody saddle-point to the calculation of a certain Liouville action on the boundary (using previous work by \cite{TZ}), the explicit form of the Liouville field for a given surface remained unknown. These explicit fields and their actions are required to determine the phase structure of dominant bulk saddles as a function of the boundary moduli. In this paper we complete this programme by calculating the Liouville field numerically.

We will also compare the action for the handlebody solutions constructed in this way to the action for a class of non-handlebodies obtained as a quotient of a Euclidean wormhole with two Riemann surface boundaries.  The action for these non-handlebodies can be easily calculated analytically. 

Recently, these higher genus partition functions have seen renewed interest from two different points of view. The states defined by integrating over half of a higher genus surface have been studied by \cite{Balasubramanian:2014hda,Maxfield:2014kra,Marolf:2015vma} (following \cite{Skenderis:2009ju}) as examples of systems with interesting entanglement structure. The slice where we cut the surface in half has several disconnected components, so the state lives in several copies of the Hilbert space of a CFT on a circle, and its entanglement can be related to connectedness of the bulk solution, extending the basic relation between the thermofield double state and eternal black holes \cite{Maldacena:2001kr}. Another motivation comes from the consideration of R\'enyi entropies: these are defined as 
\begin{equation}
S_n = \frac{1}{1-n} \ln \mbox{tr}(\rho^n)\;, 
\end{equation}
and generalise the von Neumann entropy $S = - \mbox{tr}(\rho \ln \rho)$, which can be obtained as the limit $n \to 1$.  When $\rho$ is the reduced density matrix of some spatial region $A$, obtained by tracing out the complementary region in the vacuum state (for example), the von Neumann entropy is an entanglement entropy, and the R\'enyi entropies for integer $n$ are calculated by a path integral on a `replicated surface' (with singular metric), formed by joining together $n$ copies of the sphere across the region $A$ (reviewed in \cite{Calabrese:2009qy}). This gives a genus $g =n-1 $ Riemann surface which lives in a subspace of the moduli space with a $\mathbb Z_n$ symmetry. The CFT partition functions on these surfaces were studied holographically in \cite{Faulkner:2013yia}, matching a CFT calculation \cite{Hartman:2013mia}. Since we work in a different conformal frame, our approach as implemented here does not directly compute these R\'enyi entropies. Nevertheless, we still get indirect information, and our method could be adapted if desired.

Motivated by these developments, we revisit the problem of calculating the action for the bulk handlebody saddles where the boundary is a general Riemann surface. We introduce a numerical approach for determining the Liouville field using finite element methods. We can then use this numerical solution to calculate the action for various different competing bulk saddles as a function of the moduli, allowing us to map out the phase diagram for an arbitrary Riemann surface. 

We will explicitly carry out the calculation in the genus two case in a two-dimensional subspace of the moduli space. This is sufficiently rich to exhibit an interesting phase structure, with regions where all three different phases considered in \cite{Balasubramanian:2014hda} become dominant. We map out this phase structure, confirming the qualitative expectations from previous works, and showing in particular that the non-handlebody phases are always subdominant to the handlebodies. The subspace also contains a line with a $\ZZ_3$ symmetry, corresponding to Riemann surfaces that would arise in the calculation of the R\'enyi entropy $S_3$ for a pair of disjoint intervals. We show that along this line the handlebodies respecting the $\ZZ_3$ symmetry always dominate over the symmetry breaking handlebodies we consider, though not all discrete symmetries are necessarily respected by the bulk.

We also obtain explicit results for partition functions of general CFTs in pinching limits as a cycle of a Riemann surface shrinks to zero size, which we compare to our numerical results.


In the next section, we review the work of \cite{Krasnov:2000zq,Krasnov:2003ye}, relating the action of bulk saddles to the Liouville action, which sets the stage for our work. In section \ref{surf}, we discuss the two parameter family of genus two surfaces we study to illustrate our approach, along with various analytic results. We then briefly discuss calculation of CFT partition functions in pinching limits in section \ref{sec:pinching}, and then in section \ref{num}, we explain the numerical approach we use to find the Liouville field for a given Riemann surface. In section \ref{res}, we present the results for genus two surfaces, before concluding with a discussion of the results and future work.

\section{Preliminaries}
\label{rev} 

We wish to calculate the partition function $\z$ on a compact Riemann surface $\rs$ of genus $g$ in a holographic CFT, with Einstein gravity dual\footnote{Pure Einstein gravity is a universal sector of many different holographic CFTs, all captured by our calculation. We implicitly assume that the other fields play no r\^ole in the semiclassical limit (for example, that there is no instability leading to condensation of some scalar), though the full bulk field contents are relevant at first subleading order in $c$, contributing to the one-loop determinant.}, at large central charge. In the semi-classical saddle point approximation we consider, the partition function will be given by the action $I$ of the dominant bulk classical solution, that is, the one of least action: $\z \approx e^{-I}$. As we change the moduli of the Riemann surface, the dominant bulk saddle can change, leading to first order transitions between different phases of the partition function. We compute the leading order part ($\propto c$) of the action; the order $c^0$ part can be computed from a bulk one-loop determinant \cite{Giombi:2008vd,David:2009xg}, it is sometimes possible to go to higher order in $c$ \cite{Chen:2013kpa,Chen:2013dxa,Headrick:2015gba}, and subdominant saddle points are expected to provide instanton contributions as in the genus one case \cite{Dijkgraaf:2000fq,Manschot:2007ha,Witten:2007kt,Maloney:2007ud,Keller:2014xba}, but we will say no more about such quantum corrections here.

In this section we collect the preliminary information we require. We first explain the Schottky representation of $\rs$ we use, describe how the choice of metric on $\rs$ enters the calculation, and review the construction of bulk solutions and computation of their action.

\subsection{Schottky representation of $\rs$} 

The virtue of pure three-dimensional gravity that makes our calculation tractable is that it is a locally trivial theory. This means that all Euclidean classical solutions can be constructed from quotients of Euclidean $AdS_3$, that is hyperbolic space $\HH^3$, by some subset of the isometry group\footnote{Some non-handlebody solutions come from the index two extension including orientation reversing elements.} $PSL(2,\CC)$. The boundary of $\HH^3$ is the Riemann sphere $\PP^1$, appearing as the complex plane if we adopt the upper-half space representation of $\HH^3$. The bulk isometries act on this boundary by M\"obius maps $w \mapsto \frac{aw +b}{cw + d}$. Thus, when we consider bulk solutions with boundary $\rs$, the Riemann surface $\rs$ will appear as a quotient of the Riemann sphere $\PP^1$. This is a Schottky uniformisation of $\rs$, which we now review. 

Consider $2g$ nonintersecting simple closed curves, usually circles,\footnote{If the curves can be taken to be circles, $G$ (a group to be defined shortly) will be a classical Schottky group. There exist nonclassical Schottky groups for which this is impossible, but all of the examples we consider will be classical.} $C_i,C_i'$ for $i=1,\ldots,g$ on the Riemann sphere, so that each curve divides the sphere into an `inside' and `outside', and each curve lies to the `outside' of all others.  Let the M\"obius maps $L_i$ map the interior of $C_i$ exactly into the exterior of $C_i'$. Each of the $L_i$ has repulsive and attractive fixed points $r_i$ and $a_i$, lying inside $C_i$ and $C_i'$ respectively. For a given $L_i$, these points may be mapped by conjugation to infinity and zero, after which $L_i$ is represented by a scale and rotation $w\mapsto q_i^2 w$, for some $q_i$ with $0<|q_i|<1$. This means that these $L_i$ are loxodromic elements of $PSL(2,\CC)$ (as opposed to elliptic, for which $|q|=1$, or parabolic, for which the fixed points coincide).

The group $G$ generated by the $L_i$ is a Schottky group. The set of all the $a_i,r_i,q_i^2$ can be used to parameterise the space of possible Schottky groups. After using the overall $SL(2,\CC)$ to normalise the group to some convention, say by setting $a_1=0,r_1=\infty,a_2=1$, the set of all parameters for which the $L_i$ generate a Schottky group is a domain in $\CC^{3g-3}$, known as Schottky space, in one-to-one correspondence with (normalised) marked Schottky groups, that is to say Schottky groups with a distinguished set of generators $L_1,\ldots,L_g$.

After we have removed the closure of the set of fixed points of $G$, to leave the `domain of discontinuity' $\Omega(G)\subset\PP^1$, we may take the quotient by $G$. A fundamental domain for this quotient is the region $D$ exterior to all the $C_i,C_i'$, which is a sphere with $2g$ holes cut out. The action of the quotient is to identify the circles bounding $D$ in pairs $C_i,C_i'$ by $L_i$, and each such identification acts to add a handle, resulting in a surface of genus $g$. The complex structure is inherited from that of $\PP^1$, since the M\"obius maps are conformal automorphisms.

Every Riemann surface can be obtained in this way, as a quotient $\rs=G\backslash\Omega(G)$. In fact, there are many ways to do this: picking a canonical basis $\alpha_i,\beta_i$, $i=1,\ldots,g$ with $\prod_i [\alpha_i,\beta_i]=1$ for the fundamental group of $\rs$, there is a marked Schottky group with $\rs=G\backslash\Omega(G)$, where the image of $C_i$ is freely homotopic to $\alpha_i$, and some curve between $C_i$ and $C_i'$ is homotopic to $\beta_i$. Roughly speaking, the $\alpha_i$ are a choice of $g$ cycles to cut along to get a sphere with $2g$ holes. The set of possible Schottky groups uniformising $\rs$ is in correspondence with the possible normal subgroups generated by $\alpha_i$ and their conjugates, $\mathcal{N}\langle\alpha_1,\ldots,\alpha_g\rangle$. In the context of the handlebody geometries we will describe shortly, this subgroup has a geometric interpretation, as the set of cycles that are contractible in the bulk.

At genus 1, every Schottky group is conjugate to $\{w\mapsto q^{2n}w :n\in\ZZ\}$ for some $q$, which we can write as $q=e^{i\pi\tau}$ for $\tau$ in the upper half plane; then the Schottky representation is related to the more familiar description of the torus, as the plane modulo a lattice $\CC/\langle 1,\tau\rangle$, by an exponential map $w=\exp(2\pi i z)$. Taking $\tau\mapsto\tau+1$ leaves $q^2$ unchanged, so does not alter the Schottky group. General modular transformations $\tau\mapsto \frac{a\tau+b}{c\tau+d}$ in $PSL(2,\ZZ)$ give a torus with the same complex structure, but a different Schottky group, corresponding to a choice of $\alpha$ as a primitive cycle on the torus, which goes from the origin of the $z$-plane to $m+n\tau$ for some coprime integers $m,n$.

With the exception of the torus, it is a hard problem to identify when two different Schottky groups give Riemann surfaces with the same complex structure. A crucial part of our calculation is to solve this `moduli matching' problem numerically.

\subsection{The boundary metric}
\label{bmetric}

In two dimensions the partition function of a CFT depends not just on the conformal structure, but also on the metric itself, through the Weyl anomaly. The change in the partition function under a Weyl transformation $g\mapsto e^{2\phi}g$ depends on the theory in question only through its central charge, and not on other details, being determined by the Liouville functional of $\phi$:
\begin{equation}\label{confAnomaly}
	\z[e^{2\phi}g] = e^{I_L} \z[g] \;,\quad I_L = \frac{c}{24\pi} \int \dd^2 x \sqrt{g} \left( g^{ab}\partial_a\phi\partial_b\phi + R \phi \right)
\end{equation}
The anomaly means that before we compute the partition function we need to specify which metric we are putting on the Riemann surface. 

For higher genus Riemann surfaces $\rs$ (for genus $g\geq2$), there is a unique metric of constant negative curvature on $\rs$. We will choose to study the CFT with this metric with Ricci scalar $R=-2$. In the Schottky uniformisation representation, this is related to the flat metric $\dd s^2=\dd w\dd\bar{w}$ on the $w$-plane by a conformal factor,
\begin{equation} \label{smet}
\dd s^2_\rs = e^{2 \phi} \dd w\dd\bar{w} = e^{2 \phi} (\dd x^2 + \dd y^2)\;,
\end{equation}
where for $\dd s^2_\rs$ to have Ricci scalar $R=-2$, $\phi$ satisfies the Liouville equation
\begin{equation} \label{liouville} 
\nabla^2 \phi = 4\partial_w \partial_{\bar{w}}\phi = (\partial_x^2 + \partial_y^2) \phi =e^{2\phi}. 
\end{equation}
Furthermore, we must make the metric invariant under the action of the Schottky group. The flat metric on the $w$ plane does not respect this invariance, so this requirement translates into the equivariance condition for $\phi$
\begin{equation} \label{phitrans}
\phi(L(w)) = \phi(w) - \frac{1}{2} \log |L'(w)|^2,
\end{equation}
where $L$ is any element of the Schottky group.

We note in passing that the calculation of R\'enyi entropies requires a different metric, which is flat except for conical singularities at specified points, corresponding to the ends of intervals. The conformal factor which passes to that metric can be similarly calculated as a solution of the Laplace equation with point sources at the singularities, satisfying the same boundary conditions \eqref{phitrans}. We will not solve this in the current paper, but it should be tractable using similar methods to those we employ. This offers an alternative approach to that of \cite{Faulkner:2013yia}, essentially equivalent to a two dimensional electrostatics problem with nonstandard boundary conditions.

Once we choose a fundamental domain $D$ for the Schottky group, we can regard \eqref{liouville} as a PDE on $D$, with quasiperiodic boundary conditions \eqref{phitrans} relating $\phi$ on $C_i$ and $C_i'$. The uniqueness and existence of the constant curvature metric on $\rs$ implies that there is a unique solution to this boundary value problem. Our main task will be to find this solution numerically.

This representation of the constant negative curvature metric on the Riemann surface as a scalar function on the domain $D$ has been studied in the mathematical literature, with several authors providing bounds on the conformal factor $e^\phi$ (see e.g. \cite{beardon,harmelin}). Later, we will solve \eqref{liouville} numerically for the genus two case. These solutions may be of some mathematical interest independent of the application to holography. 

\subsection{Bulk saddle points}

We now discuss the bulk saddle-points which can contribute to the holographic partition function for a given Riemann surface $\rs$. The construction of $\rs$ in terms of a quotient by a Schottky group was adopted because it provides a simple representation of the handlebody solutions in the bulk. The M\"obius maps on the boundary sphere have an extension into the $\HH^3$, where they act as orientation-preserving isometries. The loxodromic maps act without fixed points in the bulk, which means that $\bulk = G\backslash \HH^3$ is a smooth hyperbolic manifold, with boundary $\rs$.

In the case that the fundamental domain $D$ on the boundary is bounded by circles, there is a simple extension to a bulk fundamental domain, the region between the $2g$ hemispheres ending on the circles $C_i$, $C_i'$ in the Poincar\'e half-space model of $\HH^3$. The quotient then identifies these hemispheres in pairs, just as it identified the circles on the boundary. From this it is easy to see that the cycles $\alpha_i$ which are the images of $C_i$ are contractible in the bulk, while the cycles $\beta_i$ which are the image of curves from $C_i$ to $C_i'$ remain non-contractible. These $g$ noncontractible cycles generate the fundamental group of the bulk, which is free on $g$ generators; in fact, $\pi_1(\bulk)$ is naturally isomorphic to the Schottky group $G$ itself.

Each free homotopy class of closed curves curves has a unique geodesic representative, the lengths of which enter the calculation of entropies of subsets of boundaries, and which often have physical interpretations as horizons. It is therefore useful to relate the parameters of the Schottky generators to these bulk geodesic lengths. This can be done algebraically \cite{Maxfield:2014kra}, with the result that the geodesic related to the Schottky group element $g\in G \subseteq SL(2,\CC)$ has length
\begin{equation}\label{eq:geoLengths}
	\lambda(g) = \cosh^{-1}\left[\left|\frac{\tr g}{2}\right|^2 + \left|\left(\frac{\tr g}{2}\right)^2-1\right| \right]
\end{equation}
which simplifies to $\lambda=2\cosh^{-1}\frac{\tr g}{2}$ when $\tr g > 2$, as it will be in all cases we consider.

As observed above, there are an infinite number of handlebodies bounded by any particular Riemann surface, even for genus one. Roughly speaking, there is one handlebody for each choice of $g$ non-intersecting, homologically independent cycles to fill in. While it is impossible to check all the possibilities, one can make the problem more tractable by restricting attention to a subspace of the moduli space with more symmetry. With a symmetric Riemann surface, one usually expects the path integral to be dominated by a bulk saddle that preserves as much of the symmetry as possible, which will allow us to restrict investigation to finitely many geometries. Having said this, we will also check that certain saddles breaking symmetries are subdominant to a saddle preserving them.

These handlebodies are the most intensively studied bulk solutions, as they are expected to dominate the partition function, but these are not the only solutions. One simple class is obtained from the construction of Euclidean wormhole solutions with two Riemann surface boundaries by taking the bulk metric to be \cite{Maldacena:2004rf} 
\begin{equation} \label{wh} 
\dd s^2 = \dd\chi^2 + \cosh^2 \chi \; \dd s^2_\rs\;.
\end{equation}
where $\dd s^2_\rs$ is the constant negative curvature metric on $\rs$. This is a solution in the bulk for any $\rs$. If $\rs$ were the hyperbolic plane $\HH^2$, this is just the metric of $\HH^3$ in hyperbolic slicing; we can obtain any surface of genus $g\geq2$ by quotienting this $\HH^2$ by a Fuchsian group. To obtain a single-boundary solution, we can quotient this wormhole solution by a $\mathbb Z_2$ symmetry which combines $\chi \mapsto -\chi$ with an involution of $\rs$. This will provide a smooth bulk saddle with a single boundary $\rs$ if the involution acts freely, but otherwise it will have orbifold singularities at $\chi =0$ where the involution has fixed points. These are non-handlebody solutions, as none of the non-contractible cycles of $\rs$ become contractible in the bulk. Their role holographically is somewhat mysterious, as is that of the two-boundary Euclidean wormholes, and it has been conjectured \cite{Yin:2007at} that they are always subdominant relative to the handlebody solutions. We will confirm this numerically in the subspace of the genus two moduli space we consider.

\subsection{Bulk action}

To compare the different bulk saddle-points, we need to compare their actions. The evaluation of the bulk action for the handlebody solutions was discussed in detail in \cite{Krasnov:2000zq}. The gravity action we are evaluating, including all boundary contributions and counterterms, is
\begin{equation} \label{action}
	I[g] = \frac{-1}{16\pi G} \left[ \int_{\mathcal M} \dd^3x \sqrt{g} (R+ 2) +  2 \int_{\partial \mathcal M} \dd^2x \sqrt{\gamma} (\kappa - 1) -2\pi\chi\left(1+\log\frac{4 R_0^2}{\epsilon^2}\right) \right]\;,
\end{equation}
where we have set the bulk AdS scale to unity, and $\chi$ is the Euler character of the boundary. Using a Fefferman-Graham coordinate system near the boundary, with
\begin{equation}
ds^2 = \frac{\dd z^2 + \dd s^2_\rs}{z^2} + \cdots, 
\end{equation}
the bulk is cut off at the surface $z=\epsilon$, on which $\gamma$ is the induced metric, and $\kappa$ the mean curvature. The unfixed parameter $R_0$ is the radius of the sphere for which the partition function is unity, which just sets the overall normalisation for all partition functions. 

On-shell, the action evaluates to
\begin{equation} \label{bonshell}
	I = -\frac{c}{24\pi}\left[ -4V_{\mathcal M}+2A_{\partial \mathcal M} - 4\pi(g-1)\left(1-\log\frac{4R_0^2}{\epsilon^2}\right) \right]\;,
\end{equation}
where $V_{\mathcal M}$ and $A_{\partial \mathcal M}$ are the volume of the spacetime up to the cutoff and the area of its boundary\footnote{Note that \cite{Krasnov:2000zq} missed a finite term coming from a subleading part of $\kappa$, which gives a moduli independent piece proportional to $\chi$.}. The key result of \cite{Krasnov:2000zq} is that writing the boundary metric $\dd s^2_\rs$ in terms of $\phi$ as in \eqref{smet}, we can relate the action of the bulk geometry to an action for the boundary Liouville field $\phi$, 
\begin{equation}
	I = -\frac{c}{24\pi}\left[I_{TZ}[\phi] - A - 4\pi(g-1) \left(1-\log 4R_0^2\right) \right]\;.
\end{equation}
The term $A$ is the area of the boundary, which depends on the boundary metric; it is cancelled by an identical term appearing in $I_{TZ}$. The nontrivial dependence is given by $I_{TZ}$, which is the Liouville action of Takhtajan and Zograf \cite{TZ},
\begin{align} \label{TZact}
	I_{TZ}[\phi] =& \int_D \frac{i}{2} \dd w \wedge \dd\bar{w} \;  \left(4\partial \phi \bar{\partial} \phi +e^{2\phi}\right) \\
	  &+\sum_{k=2}^{g} \int_{C_k} \left(2\phi -\frac{1}{2}\log |L'_k|^2 -\log|c_k|^2\right) \frac{i}{2}\left(\frac{L''_k}{L'_k} d w - \frac{\bar{L}''_k}{\bar{L}'_k} d\bar{w} \right)\;. \nonumber
\end{align}
The boundary terms are such that the value of the action is independent of the choice of fundamental region $D$ used to evaluate it. This form assumes that $L_1$ has a fixed point at infinity, so in particular, $D$ is a bounded domain of the complex plane.\footnote{This restriction can be relaxed, either allowing $L_1$ to be more general and including $k=1$ in the sum, and if $D$ is unbounded with the addition of a regulating circle at large radius, on which there is an additional boundary term \eqref{unboundedD}.} The functional $I_{TZ}$ is stationary exactly on the constant negative curvature metrics, solving the Liouville equation \eqref{liouville}.

Thus, the barrier to evaluating the bulk action for the handlebody phases is simply lack of knowledge of the Liouville field that corresponds to the desired metric on $\rs$, which for our choice is the solution of the Liouville equation \eqref{liouville}. We will show in section \ref{num} how to solve this equation numerically; it is then straightforward to perform the above integrals on the numerical solution to obtain the action.

We will also compare the values of the action for the handlebody phases to the action for the non-handlebody solution \eqref{wh}, which is much easier to evaluate because of the warped product form of the metric. Evaluating the action \eqref{bonshell} for the metric \eqref{wh} with $\chi \geq 0$, the $\chi$ integral is independent from the integral over $\rs$, so the whole action is proportional to the area of $X$ in the constant negative curvature metric, which is $4\pi(g-1)$ by the Gauss-Bonnet theorem. The divergent terms cancel, to leave us with
\begin{equation}
I  = -\frac{c}{3} (g-1) \log R_0
\end{equation}
for the non-handlebody solution. 

For our results below, we will normalise $\z$ by choosing $R_0=1$, so in particular the action for the non-handlebody is zero.

\section{A two-parameter family of surfaces}
\label{surf}

Our discussion so far has been completely general; it could be applied to the calculation of the bulk action for any Riemann surface $\rs$ at any point in its moduli space. To illustrate our numerical methods, we will explictly compute $\phi$ for a simple family of genus two surfaces: we consider a two real-dimensional subspace of the full three complex-dimensional moduli space of genus two surfaces. This is a subspace of the three real-dimensional space of surfaces with a reflection symmetry considered in \cite{Balasubramanian:2014hda}, describing multiboundary wormholes. This same subspace also arises as a subspace of the set of genus two surfaces which can be interpreted as states with a single boundary. In addition to the physical interest of these interpetations, the symmetry of $\rs$ in this subspace allows us to restrict consideration to bulk saddle points which preserve this symmetry, so we will be able to completely map out the relevant phases. In this section, we describe this subspace in more detail, and identify the bulk saddle points which preserve the symmetry and explain their physical interpretations.

\subsection{Reflection symmetric surfaces}

We first restrict to a class of Riemann surfaces with a reflection symmetry. Mathematically, this can be described as a \emph{real structure}, which is a notion of complex conjugation on the surface. This is defined as an involutive anticonformal automorphism $\sigma$. Further, we will only be interested in the case where the fixed point set of $\sigma$ (which is known as the real part, and always consists of a finite number $k$ of simple closed curves) divides the surface into two components. Real surfaces $(\rs,\sigma)$ can be classified by their `topological type' $(g,k,\epsilon)$, where $g$ is the genus, $k$ the number of components of the real part as above, and $\epsilon$ determines whether the fixed point set of $\sigma$ splits the surface in two ($\epsilon=0$, the case we require) or leaves it connected ($\epsilon=1$). A real surface $(\rs,\sigma)$ of type $(g,k,0)$ can be built by taking a Riemann surface $Y$ with $k$ boundaries, and sewing it to a reflected copy of itself along the boundaries, a construction called the Schottky double. Then  $\sigma$ is the involution that simply swaps these two copies. The restriction to surfaces with a reflection symmetry is physically motivated by the interpretation of these surfaces as defining a natural set of states on the tensor product $\mathcal{H}^{\otimes k}$ of $k$ copies of the CFT. The wavefunction evaluated on any field configuration is defined by the path integral on $Y$ with the given field configuration as boundary conditions. 

At genus zero, this is just the state-operator correspondence definition of the vacuum, and for genus one, where the reflection symmetry restricts us to the rectangular torus ($\tau=i\beta/2\pi$ pure imaginary), this construction gives the thermofield double state.

 For a holographic CFT, these states have geometric duals \cite{Skenderis:2009ju} which are black holes with $k\geq1$ asymptotic boundaries, and arbitrary topology behind the horizon \cite{Aminneborg:1997pz,Brill:1995jv,Brill:1998pr}. Depending on the moduli of $\rs$, there may be several different possible bulk duals, with first order phase transitions between them as we vary the moduli. The dominant solution is expected to respect the involution symmetry $\sigma$, which extends into the bulk as a time-reflection symmetry. This fixes a $t=0$ slice $\tzero$ in the bulk, where the quantum state of the bulk fields is defined by a Hartle-Hawking procedure, integrating over configurations on half of the Euclidean bulk, just as in the definition of the boundary state. Because $SL(2,\RR)$ acts on the equatorial slice of $\HH^3$ in precisely the same way as it acts on the Northern and Southern hemispheres of the boundary, there is always a phase in which $\tzero$ is conformal to $Y$, but there are also phases in which $\tzero$ is disconnected, so the state in different copies of the CFT may be entangled only through the quantum state of the bulk fields, and not at leading order in the large $c$ semiclassical limit. Since $\tzero$ has vanishing extrinsic curvature, it provides a good Cauchy slice from which to classically evolve forwards and backwards in Lorentzian time.
 
 This general structure can be illustrated by considering the familiar case of genus one, where everything can be explicitly computed. The rectangular torus (type $(1,1,0)$) has two saddle points respecting the reflection symmetry, filling in either the thermal Euclidean time circle, or the spatial circle. In the first case, the bulk $t=0$ slice is a cylinder connecting the two boundary components, which is the Einstein-Rosen bridge of the BTZ black hole, and in the second case $\tzero$ is two disconnected copies of $\HH^2$, the $t=0$ slice of pure $AdS$. The leading classical contribution to the torus partition function always comes from one of these solutions, which exchange dominance at the square torus $\tau=i$, the $AdS_3$ version of the Hawking-Page phase transition.
 
 \subsection{Genus two}
  
We now describe the moduli space of real surfaces at genus two. 
There are two topologically distinct types of surfaces admitting a time-reflection symmetry, each with three real moduli. The first, with $(X,\sigma)$ of type $(2,3,0)$, studied in \cite{Balasubramanian:2014hda}, is the Schottky double of a pair of pants $Y$, describing an entangled state on three copies of the CFT. The moduli space can be parameterised by the lengths (in the canonical constant curvature metric on $\rs$) of the three circles fixed by $\sigma$,  $L_a$, $a=1,2,3$. There are three kinds of handlebody saddles which preserve the $\mathbb Z_2$ symmetry, and thus are expected to dominate the path integral, distinguished by the topology of the time reflection invariant surface $\tzero$ in the bulk. For the first, $\tzero$ is a pair of pants, conformal to $Y$, corresponding to a connected wormhole geometry. For the second, $\tzero$ is  three discs, corresponding to three disconnected copies of $AdS$. Finally it can be a disc plus a cylinder, with three different versions depending on which two boundaries the cylinder joins, corresponding to one copy of $AdS$ and one of BTZ.
 
The second possibility is type $(2,1,0)$, so $\sigma$ fixes just one curve, and $Y$ is a torus with a single boundary. These surfaces define some excited pure state in a single copy of the CFT, once again with three real moduli. For one possible parameterisation of this space, choose a dual pair of cycles on the torus, and give their lengths in the canonical metric on $\rs$, as well as the angle at which they meet. Another possibility is the Fenchel-Nielsen coordinates, building the bordered torus from a pair of pants, with two cuffs the same length, joining those cuffs with some twist. There are now infinitely many different handlebodies respecting the time reflection symmetry. One has the $t=0$ slice $\tzero$ equal to $Y$:  this is a single-exterior black hole with a torus hidden behind the horizon. The other phases have $\tzero$ simply a disc, and are classically dual to pure $AdS$, with the quantum state of the bulk fields depending on which of the possible handlebodies is dominant.
 
These two three-parameter subspaces of Riemann surfaces admitting different real structures in fact intersect in a two-dimensional family that admits both types of time-reflection symmetry. From the point of view of the three boundary wormhole, this is the restriction to the subspace where two of the boundaries have the same size, $L_1=L_2$. There is then a reflection symmetry that swaps these two cycles; this is the symmetry that is interpreted as time reflection for the torus wormhole. For the torus, the additional symmetry arises when we demand that there is a dual pair of cycles meeting at right angles, so we have a rectangular torus. We will analyse this two-dimensional subspace. 

There are three possible bulk geometries respecting both of these symmetries. For the three boundary wormhole interpretation, they correspond to a wormhole connecting boundaries of equal size with a disconnected $AdS$ for the third, the totally connected wormhole joining all three boundaries, and a totally disconnected geometry with three copies of $AdS$. In the alternative interpretation, these phases correspond to the single exterior black hole phase, and two different pure $AdS$ phases, where one or other of the cycles of the torus is contractible in the Euclidean bulk.

Solutions in this subspace in fact all have a third reflection symmetry, with total conformal and anticonformal symmetry group $\ZZ_2 \times \ZZ_2 \times \ZZ_2$. This additional symmetry can be understood from the fact that every genus 2 surface is hyperelliptic, so it can be represented as an algebraic curve $y^2=p(x)$ for some polynomial $p$ of degree five or six; every such surface admits the hyperelliptic involution $y\mapsto -y$. When this commutes with the real structure $\sigma$, the combination of $\sigma$ with the hyperelliptic involution provides a second inequivalent real structure. For a surface formed by gluing two copies of a pair of pants, this second real structure can be thought of as reflection in the plane splitting the pants in two along the seams. If we call the cuffs of the pants the A-cycles, fixed by the original time reflection, the second real structure fixes a dual set of B-cycles. The surface thus may be reinterpreted by swapping the names of these cycles, and in this interpretation represents a different point in the moduli space of real surfaces of type $(2,3,0)$. This is much the same as the case for a rectangular torus, where there is a spatial reflection symmetry in addition to the time reflection symmetry, so swapping the r\^oles of space and time gives a different interpretation of the same surface. Any handlebody solution can also be interpreted from these two points of view, so a Euclidean geometry corresponding to a connected phase can be reinterpreted as the disconnected phase for a different point in moduli space.

The full order 8 symmetry group is manifest in the Schottky groups and fundamental domains $D$ we choose, as reflections and inversions. This is be useful for increasing the numerical efficiency, since it allows us to solve the Liouville equation in a region $\tilde{D}$ one eighth the size of the full fundamental domain, inferring the values elsewhere by the equivariance of $\phi$ under the symmetries. Expressions for the action in the form of integrals only over $\tilde{D}$ and its boundaries are given in appendix \ref{integrals}. The calculations in the reduced domain agree to numerical accuracy with those done in the full domain with the original action \eqref{TZact}, but are more computationally efficient. The fundamental domains for the Schottky groups we use are shown in figure \ref{SchottkyDomains}, and described in more detail in the appendix.

\begin{figure}[htbp]
	\centering
	\includegraphics[width=.55\textwidth]{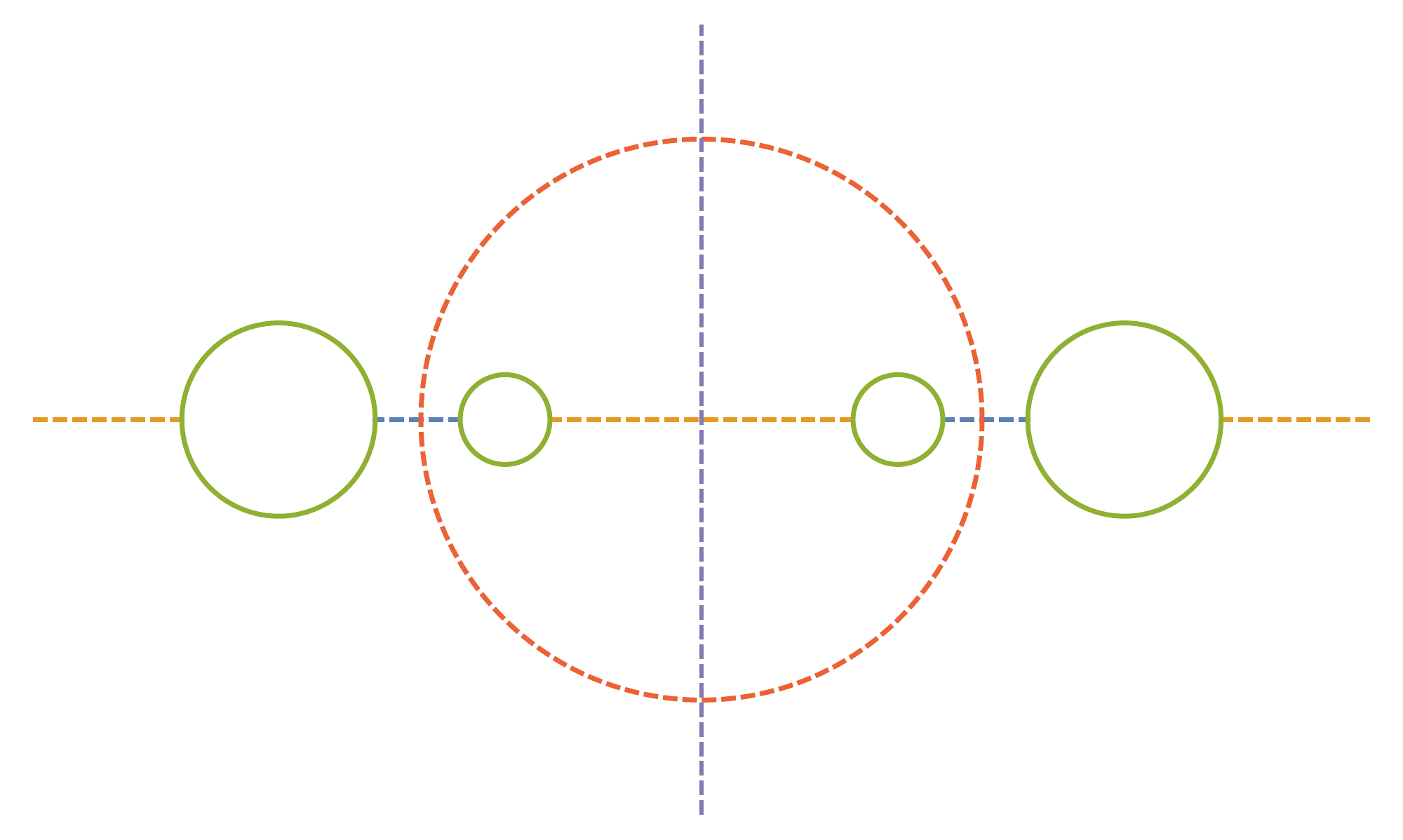}
	\includegraphics[width=.33\textwidth]{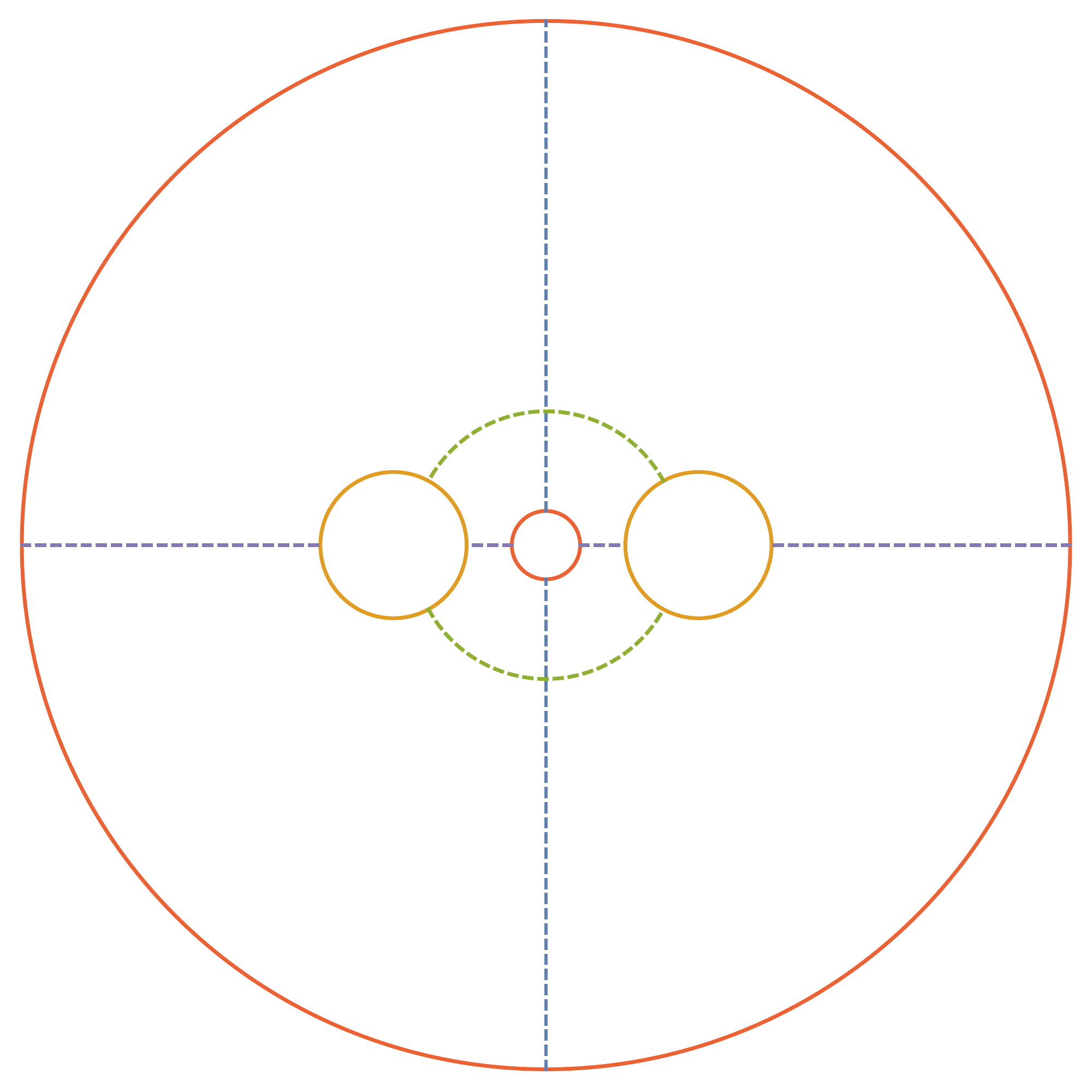}
	\\
	\includegraphics[width=.9\textwidth]{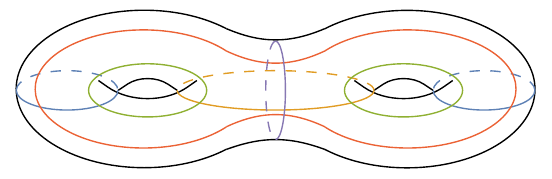}
	\caption{The choice of fundamental domains $D$ we use for the relevant Schottky groups $G$, corresponding in the three boundary wormhole to the connected phase (left) and partially connected (right), and a cartoon of the surface we consider with various relevant cycles marked. The generators for the Schottky group for the left domain identify the two green circles on the right with one another, and similarly the two on the left; for the right domain they identify the concentric orange circles, and the yellow circles. The disconnected phase uses the left domain, but swaps the interpretations of A- and B-cycles. The three involutions which generate the symmetries of the surface can be visualised as reflections in the three coordinate axes. Reflection in the horizontal plane fixes the blue and yellow A-cycles, and is interpreted as the time-reversal for the three boundary wormhole, implemented as reflection in the horizontal axis for the left domain, and the vertical axis for the right. Reflection in the plane of the page, fixing green and orange B-cycles, is implemented as inversion in the unit circle for both domains. Finally, the reflection in the vertical plane fixing the purple `waist', interpreted as time reversal for the torus wormhole, is represented by reflection in vertical and horizontal axes for left and right domains respectively. The hyperelliptic involution is the half rotation around the horizontal axis, fixing the six points where A- and B-cycles intersect.}
	\label{SchottkyDomains}
\end{figure}

 \subsection{Enhanced symmetries}
 \label{symm}
 
Our Riemann surfaces  generically have a symmetry group of order 8, but there are subspaces where the symmetry is further enhanced. There are two distinct one-parameter families of surfaces in our moduli space for which such enhancement occurs, as follows from the classification \cite{cirre2001complex}, which help to understand some aspects of the phase diagram analytically. These two families meet at a distinguished point in moduli space, a very highly symmetric Riemann surface; this is the genus 2 analogue of the square torus $\tau=i$. We can identify explicitly the hyperelliptic curve corresponding to this surface, and calculate its cycle lengths analytically, which will act as a check on the numerical results. 
 
 The first possible enhancement occurs when all three A-cycles (and hence also all three B-cycles) have the same length, corresponding to three-boundary wormholes with all horizons the same size. The full conformal automorphism group for these surfaces is $D_6$, the dihedral group of order 12, along with the anticonformal time reflection commuting with these.  Here we can explicitly check that the pair of pants and thermal AdS saddles, which preserve the $\mathbb Z_3$ symmetry, always dominate over the disc plus annulus saddles, which break it. This is also the same family of surfaces relevant for computation of the third R\'enyi entropy of two disjoint intervals, parametrised by the cross ratio of the intervals' endpoints, so we are also verifying replica symmetry.

The other symmetry enhancement is an additional $\ZZ_2$, enhancing the conformal automorphism group to $D_4$. From the torus wormhole point of view, this occurs when the two cycles of the torus behind the horizon have equal size. The symmetry exchanges the A-cycles and B-cycles, which have matching lengths for these surfaces. As a consequence, the handlebodies for disconnected and connected phases, which do not respect this symmetry but are exchanged under it, are degenerate. This indicates that the symmetry enhancement is at the boundary separating these two phases. It is possible to analytically calculate the relationships between cycle lengths at this boundary, as we will describe briefly below. However, because the relevant Schottky groups do not respect this symmetry, this information is not helpful for finding the Schottky parameters, and hence the spacetime parameters such as horizon lengths, at the phase transition, which will instead be determined from the numerics. It is worth noting here that the partially connected phase does respect the full symmetry group along this line, but is still subdominant for some values of moduli, so this is an example where the bulk spontaneously breaks a discrete symmetry.

Now these two families of enhanced symmetry meet at a distinguished point in the moduli space of curves where all six A-cycles and B-cycles are all the same length. This is the genus 2 analogue of the point $\tau=i$ in the torus moduli space, where the horizontal and vertical cycles are the same length\footnote{This is occasionally referred to as the Bolza surface, though this name is more usually given to the maximally symmetric genus 2 surface, with twice as many automorphisms, but only real structures of type $(2,1,1)$ or $(2,2,1)$, analogous to the torus with $\tau=e^{i\pi/3}$.}. From the classification \cite{cirre2001complex}, there is a unique surface with the appropriate enhancement of symmetries, with a conformal automorphism group of order 24, corresponding to the hyperelliptic curve $y^2=x^6-1$
. For this surface, it is possible to explicitly work out the lengths of geodesic cycles on the constant curvature metric, which gives us a check on the results.

\begin{figure}[htbp]
 \centering
	\includegraphics[width=.45\textwidth]{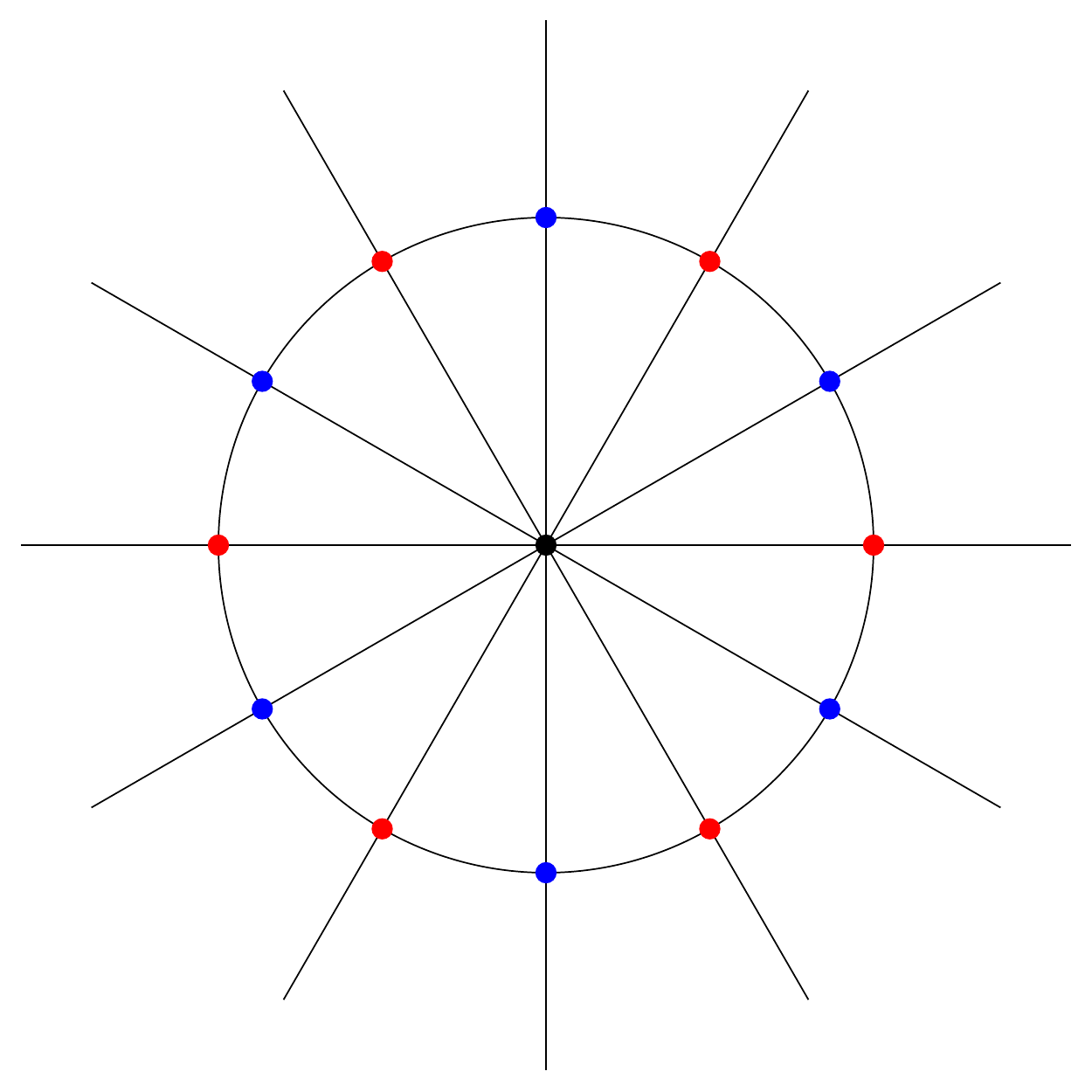}
	\includegraphics[width=.45\textwidth]{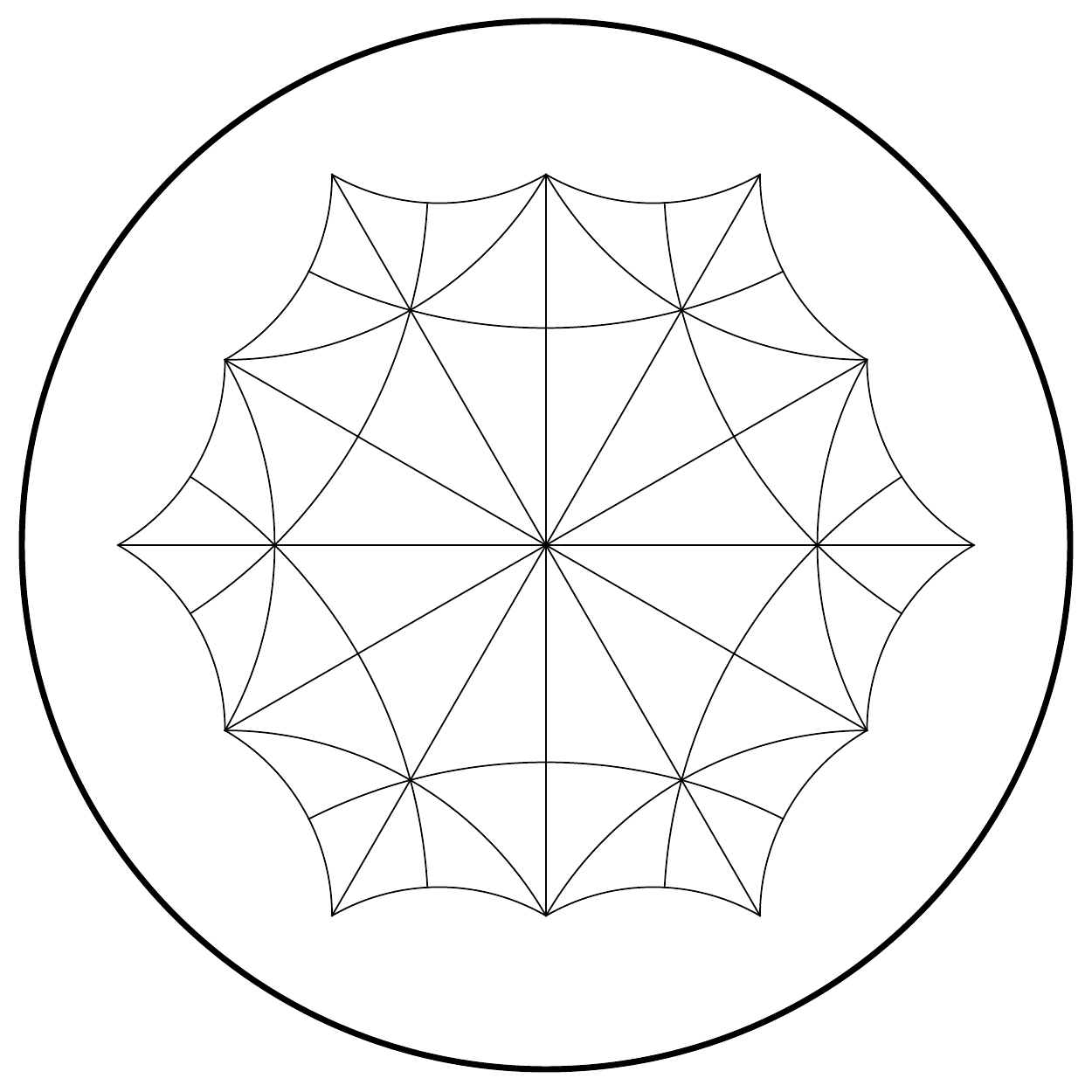}
\caption{The maximally symmetric real genus 2 Riemann surface. The left plot shows a tessellation of the complex $x$-plane respecting the symmetries of the curve $y^2=x^6-1$. The symmetries acting on the $x$-plane include the obvious reflections and rotations, as well as inversion in the unit circle and the hyperelliptic involution $y\mapsto -y$ exchanging sheets. The red points are the sixth roots of unity, where the plane is ramified. Each triangle has a vertex at either $0$ or $\infty$ on the $x$-plane, where 12 triangles meet, so the angle is $\frac{\pi}{6}$, and a vertex at $e^{(2k+1)i\pi/6}$ for some integer $k$, where 4 triangles meet with right angles (the blue points in the figure), so the angle is $\pi/2$. Finally, there is a vertex at a sixth root of unity (the red points), which is a ramification point for the hyperelliptic curve, so there are in fact 8 triangles meeting at this point, four on each sheet, with angles $\pi/4$.  The right diagram shows a fundamental domain of the hyperbolic plane for the surface, which can be represented as a quotient of $\mathbb{H}^2$. The surface is given by an appropriate identification of edges of this domain. \label{symmetricSurf}}
\end{figure}

To do this, notice that the symmetries allow us to identify a tessellation of the curve by identical hyperbolic triangles. We can identify their edges on the complex $x$-plane, since they all lie on fixed points of some symmetry, as shown in figure \ref{symmetricSurf}. These triangles all have corner angles $\pi/2$, $\pi/4$ and $\pi/6$, and a little hyperbolic trigonometry gives the side lengths, which are $\frac{1}{2}\cosh^{-1}5$, $\frac{1}{2}\cosh^{-1}3$, and $\frac{1}{2}\cosh^{-1}2$. Now the cycles on the $t=0$ slice of the three boundary wormhole, and the corresponding dual cycles, which we measure in the numerics, correspond to arcs of the unit circle between two of the sixth roots of unity on the $x$-plane, returning by the same arc on the other sheet. These contain four of the short edges of the triangle, so are of length $2\cosh^{-1}2$. We will verify numerically that this is the cycle length at the point where all six A- and B-cycles have the same length.

With a little more effort, we can make progress in calculating cycle lengths more generally. To do this, note that if we cut our surface along all geodesics fixed by symmetry, that is all the curves marked in figure \ref{SchottkyDomains}, we decompose it into eight pentagons, with right angles at every vertex. With the constant curvature metric, these are right angle hyperbolic pentagons, with side lengths $\frac{1}{2}\ell_{1,2}^A,\frac{1}{2}\ell_{1,2}^B,\frac{1}{4}\ell_{3}^A,\frac{1}{4}\ell_{3}^A,\frac{1}{4}\ell_{w}$, where $\ell_a^{A,B}$ is the lengths of the $a$th A- or B-cycle, and $\ell_{w}$ the length of the `waist' that divides the surface in two. But there is a two-dimensional space of such pentagons, so specifying two side lengths uniquely determines the others, and in particular, the B-cycle lengths can be analytically determined as a function of the A-cycle lengths. As an example of this, the phase boundary between disconnected and connected phases, which happens when the A-cycle and B-cycle lengths are equal, occurs when
\begin{equation}\label{eq:conDiscTransition}
	\ell_3 = 4\coth^{-1} \sinh \left(\frac{\ell_{1,2}}{2}\right) \, .
\end{equation}
Note that this does not allow us to identify the Schottky parameters at the transition, because the handlebody does not respect the relevant symmetry.

\section{Pinching limits}\label{sec:pinching}

As a check on our numerically computed actions, we would like to compare them to any available analytic results. Some progress can be made at the edges of moduli space, where one or more cycles of the Riemann surface shrink to zero size. The contribution of the vacuum block to the partition function in this limit was studied in \cite{Headrick:2015gba}, which gave loop corrections to high orders in the semiclassical $1/c$ expansion. However, it is not so easy to obtain the classical, order $c$ contribution that we are studying, which is related to the fact that this depends on the choice of frame by the Weyl anomaly, and not just the complex structure of the Riemann surface. In this section, we explicitly compute how the order $c$ part of the partition function in the constant curvature frame behaves in a pinching limit, in terms of the length $\ell$ of the pinching cycle.

To do this, we consider cutting the Riemann surface along the short cycle, and insert a complete set of states, which equates the partition function to a sum of two-point functions on a simpler (possibly disconnected) Riemann surface; see \cite{Headrick:2015gba}, particularly the final appendix, for more details. The contribution from each state comes with $\ell$ to the power of the dimension, so the sum will be dominated by the vacuum state in the pinching limit. The vacuum is prepared by the path integral on hemispheres, which we sew to each side of the cut cycle, normalised by dividing by the partition function on a sphere of the same radius. So in the pinching limit, we find the leading order piece of the partition function by cutting the surface along the cycle, inserting hemispherical `caps', and dividing by the partition function for the sphere of that radius:
\begin{equation}
	\left. \z\left[\eqpic{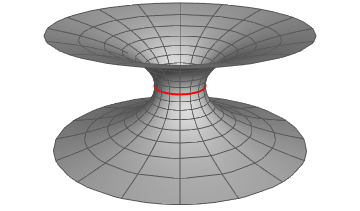}{6} \right] \sim \z\left[\eqpic{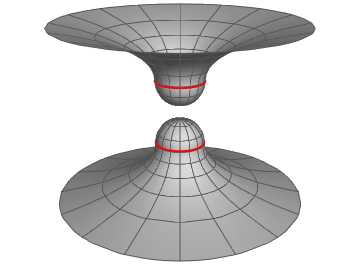}{6} \right] \middle/ \z\left[\eqpic{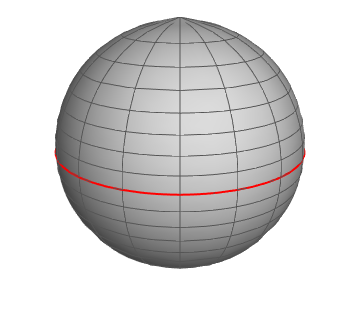}{4} \right] \right. \text{ as }\ell\to 0
\end{equation}
where $\z[\cdot]$ is the partition function of the indicated surface, and the images should be interpreted as small pieces, near the pinching cycle (marked in red), of some Riemann surface.

In any neighbourhood of the cycle topologically equivalent to a cylinder, by the uniformisation theorem the geometry is determined uniquely by the length of the cycle. The metric in can be written in the canonical form (defining $R$ by $\ell=2\pi R$)
\begin{equation}
	\dd s^2=\dd r^2+ R^2 \cosh^2 r \; \dd\theta^2
\end{equation}
with our cycle at $r=0$, valid for some range of the coordinate $r$. In the pinching limit we consider, we require that any cycle dual to the short cycle must become long, which implies that the range of $r$ in which this metric is valid becomes arbitrarily large. Once we cut along the cycle and insert a hemisphere, this metric will give the geometry for $r>0$, and then
\begin{equation}
	\dd s^2=\dd r^2+ R^2 \cos^2 \left(\frac{r}{R}\right) \dd\theta^2
\end{equation}
is the hemisphere metric for $-\frac{\pi}{2} R<r<0$.

Now there exists a conformal transformation $g\mapsto \tilde{g} = e^{2\phi}g$ on the geometry with the hemispherical cap sewn in, decaying exponentially for large $r$, which maps it to the same capped geometry, but with a different value of $R$. Since the transformation decays, in the pinching limit this will become a good approximation for a conformal map on the entire Riemann surface (cut and with hemispheres sewed on), which changes the radius of the cycle, but leaves the geometry far from it unaltered. The ratio of partition functions at the two different radii is then simply given by the Liouville action of the conformal map (along with a term coming from the transformation of the sphere). We will find this conformal map to linear order, which gives us the derivative of the partition function with respect to cycle length in the pinching limit.

There are three different regions for the conformal map, given by $r<0$, $0<r<r_0$ and $r>r_0$, where $r_0$ is the location of the geodesic cycle in the rescaled metric. In each region, we have different values of scalar curvature (denoted $\mathscr{R}$ here to distinguish from the radius of the cycle) before and after the conformal map, depending on whether the relevant point begins and ends in the constant negative curvature region or on the hemisphere. We must find $\phi$ from the differential equation prescribing the transformation of curvature
\begin{equation}
	\tilde{\mathscr{R}} = e^{-2\phi}(\mathscr{R}-2\nabla^2 \phi),
\end{equation}
solving separately for $\phi$ in each region. For $r>r_0$, we have $\mathscr{R}=\tilde{\mathscr{R}}=-2$; for $r<0$, we have $\mathscr{R}=\frac{2}{R^2}$, $\tilde{\mathscr{R}}=\frac{2}{R'^2}$; and finally for $0<r<r_0$ we have $\mathscr{R}=-2$, $\tilde{\mathscr{R}}=\frac{2}{R'^2}$. 

Picking $R'=R(1+\epsilon)$, for $\epsilon$ small, we can linearise the equation and solve (with an Ansatz independent of $\theta$). This is straightforward in the first and second regions where the curvature is unchanged or changes to linear order, but in the small region $0<r<r_0$ there is an order one change in curvature, so $\phi''(r)$ will not in fact be small.

First we solve for $r>r_0$. One boundary condition specifies that $\phi$ decays; then we find the new location $r_0$ of the constant-$r$ geodesic, and specify the remaining boundary condition by demanding that its radius be $(1+\epsilon)R$. Next, observe that $\phi$ and $\phi'$ (both of which must be continuous) are small, and converting the $0<r<r0$ region to part of a hemisphere must require $\phi''$ to be of order one. Since $r_0$ is small, so the solution need only be valid over a small region, it suffices to set $\phi$ as a quadratic in $r$ to this order of approximation. This sets the boundary conditions of $\phi$ at $r=0$, from which we may subsequently integrate to $r=-\frac{\pi}{2}R$, verifying for consistency that the result is smooth there ($\phi'=0$). The solution is
\begin{equation}
	\phi(r)\sim \epsilon\begin{cases}
		1- \tan^{-1}(\csch(r)) \sinh(r) & r>r_0=\frac{\pi}{2}\epsilon \\
		1+\frac{\pi}{2R^2}r-\frac{1}{2\epsilon}\left(1+\frac{1}{R^2}\right)r^2+\cdots & 0<r<r_0 \\
		1+\frac{\pi}{2R}\sin\left(\frac{r}{R}\right) & -\frac{\pi}{2}R<r<0
	\end{cases}
\end{equation}
Note that in the middle regime, $\phi$ and $\phi'$ are order $\epsilon$, but $\phi''$ is order one, so that the curvature can be changed at order one. It will still not contribute to the Liouville action at linear order.

We now need the conformal anomaly \eqref{confAnomaly}
\begin{equation}
	\z[e^{2\phi}g] = e^{I_L} \z[g] \;,\quad I_L = \frac{c}{24\pi} \int \dd^2 x \sqrt{g} \left( g^{ab}\partial_a\phi\partial_b\phi + \mathscr{R} \phi \right)
\end{equation}
where $\mathscr{R}$ is the scalar curvature in the metric $g$. In our case, working to linear order, only the last term in the Liouville action contributes, giving
\begin{equation}
	I_L = -\frac{c}{48} 2\pi R \epsilon + \frac{c}{6}\epsilon -\frac{c\pi\epsilon}{24 R}
\end{equation}
and the Liouville action for changing the radius of the sphere from $R$ to $(1+\epsilon) R$ is $\frac{c}{3}\epsilon$ to leading order, which cancels the constant term. The result that follows is that for the pinching limit of the surface where the calculation applies, neglecting all but the $R^{-1}$ contribution, we get
\begin{equation}
	\frac{d\log\mathcal{Z}}{dR} \sim -\frac{c\pi}{12R^2}
\end{equation}
so, integrating up, we get
\begin{equation}
	\log\mathcal{Z} \sim \frac{c}{12} \frac{\pi}{R}.
\end{equation}
The first correction to this will be independent of $R$, depending on the full details of the moduli of the punctured Riemann surface resulting from pinching the cycle to a point.

In the saddle point approximation where $\mathcal{Z}\sim e^{-I}$, this means that the dominant saddle point must behave as
\begin{equation}\label{eq:pinching}
	I \sim -\frac{c}{6} \frac{\pi^2}{\ell} \quad\text{ as } \ell\mapsto 0
\end{equation}
where we now parameterise the pinching cycle by its length $\ell=2\pi R$.

If several cycles pinch, unless they are also at a degenerate point where the cycles remain at finite distance in the limit, this calculation applies independently to all the cycles, so the result is just a sum of the various terms.

\section{Numerical solution}
\label{num}

Given a Schottky uniformisation of a Riemann surface, described as a region $D$ in the complex plane whose boundaries are circles $C_i$, $C_i'$ identified under the action of the loxodromic M\"obius maps $L_i$, our task is to find a solution $\phi$ of the Liouville equation \eqref{liouville} with the transformation properties \eqref{phitrans}, and to evaluate the action \eqref{TZact} for this solution. Here we describe a numerical method of solution.

Since the Liouville equation is nonlinear, we opt to use the Newton-Raphson algorithm.  Linearising the Liouville equation gives us the linear Newton-Raphson equation
\begin{equation}\label{newton}
(\nabla^2-2e^{2\phi})\delta\phi_{(i)}=-\left(\nabla^2\phi_{(i)}-e^{2\phi_{(i)}}\right)\;.
\end{equation}
Similarly, \eqref{phitrans} yields the quasi-periodic boundary conditions
\begin{equation}\label{newtonbc}
\delta\phi_{(i)}[L(w)]-\delta\phi_{(i)}[w]=-\left(\phi_{(i)}[L(w)]-\phi_{(i)}[w]-\frac{1}{2}\log|L'(w)|^2\right)\;.
\end{equation}
Beginning with a seed $\phi_{(0)}$, we iteratively solve the above equation and boundary conditions for $\delta\phi_{(i)}$, and then set $\phi_{(i+1)}=\phi_{(i)}+\delta\phi_{(i)}$.  The procedure continues until some success or failure condition is met\footnote{We find that the basin of attraction for this problem is large enough that using the seed $\phi_{(0)}=0$ converges for all of the solutions we have obtained.  We stop the algorithm when $||\delta\phi_{(i)}||_{\infty}<10^{-10}$.}.  For convenience, we will drop the subscripts that label the Newton iterations in the remainder of this section.

To solve the above linear differential equations numerically, we need a discretisation scheme.  This particular differential equation is simple, but lies in a complicated domain.  We therefore resort to finite elements, which are well-adapted to these kinds of problems.  

Rather than to solve the linear equation \label{newton} directly, the finite element method solves an integral form of this equation.  Multiplying the equation by a test function $\psi$ and integrating over the domain $D$, we obtain
\begin{equation}
-\int_D\nabla\psi\cdot\nabla\delta\phi-2\int_D\psi e^{2\phi}\delta\phi+\int_{\partial D}\psi \nabla_n \delta\phi = \int_D\nabla\psi\cdot\nabla\phi+\int_D\psi e^{2\phi}\phi-\int_{\partial D}\psi \nabla_n\phi
\end{equation}
where we have performed an integration by parts on the second-derivative terms, $\nabla_n$ is an outward-pointing normal derivative, and the integrals are taken using the unsigned Euclidean measure. We can now use the boundary condition \eqref{newtonbc} to rewrite the boundary terms.  In our case where $D$ has been reduced by using the involution $w\rightarrow 1/w$ symmetry, this results in 
\begin{equation}\label{integraleq}
-\int_D\nabla\psi\cdot\nabla\delta\phi-2\int_D\psi e^{2\phi}\delta\phi = \int_D\nabla\psi\cdot\nabla\phi+\int_D\psi e^{2\phi}\phi+\sum_{C_i}\int_{C_i\cap D}\psi\frac{\sigma(C_i)}{R(C_i)}\;,
\end{equation}
where $R(C_i)$ is the radius of the circle $C_i$, and $\sigma(C_i)=-1$ if the domain lies within the circle, and $\sigma(C_i)=+1$ if the domain lies outside the circle\footnote{Note that using the involution symmetry has replaced the quasi-periodic boundary boundary conditions with simple boundary integrals. Without such a symmetry, one would require some means of imposing the quasi-periodic boundary condition more explicitly, such as a master/slave method that replaces degrees of freedom on $C_i'$ with those on $C_i$.}. The domain can be further reduced by lines of reflection symmetry without changing these boundary terms.  The lines themselves have no boundary contribution, which is equivalent to a Neumann boundary condition.  

At each Newton iteration, we are given $\phi$, and the aim is to find a $\delta\phi$ such that the integral equation \eqref{integraleq} is valid for any $\psi$ in some space of functions.  For us, this will be a Sobolev space of piecewise continuous second-order polynomials.  We will also approximate $\phi$ and $\delta \phi$ using functions within this Sobolev space.

\begin{figure}[htbp]
 \centering
	\includegraphics[width=.35\textwidth]{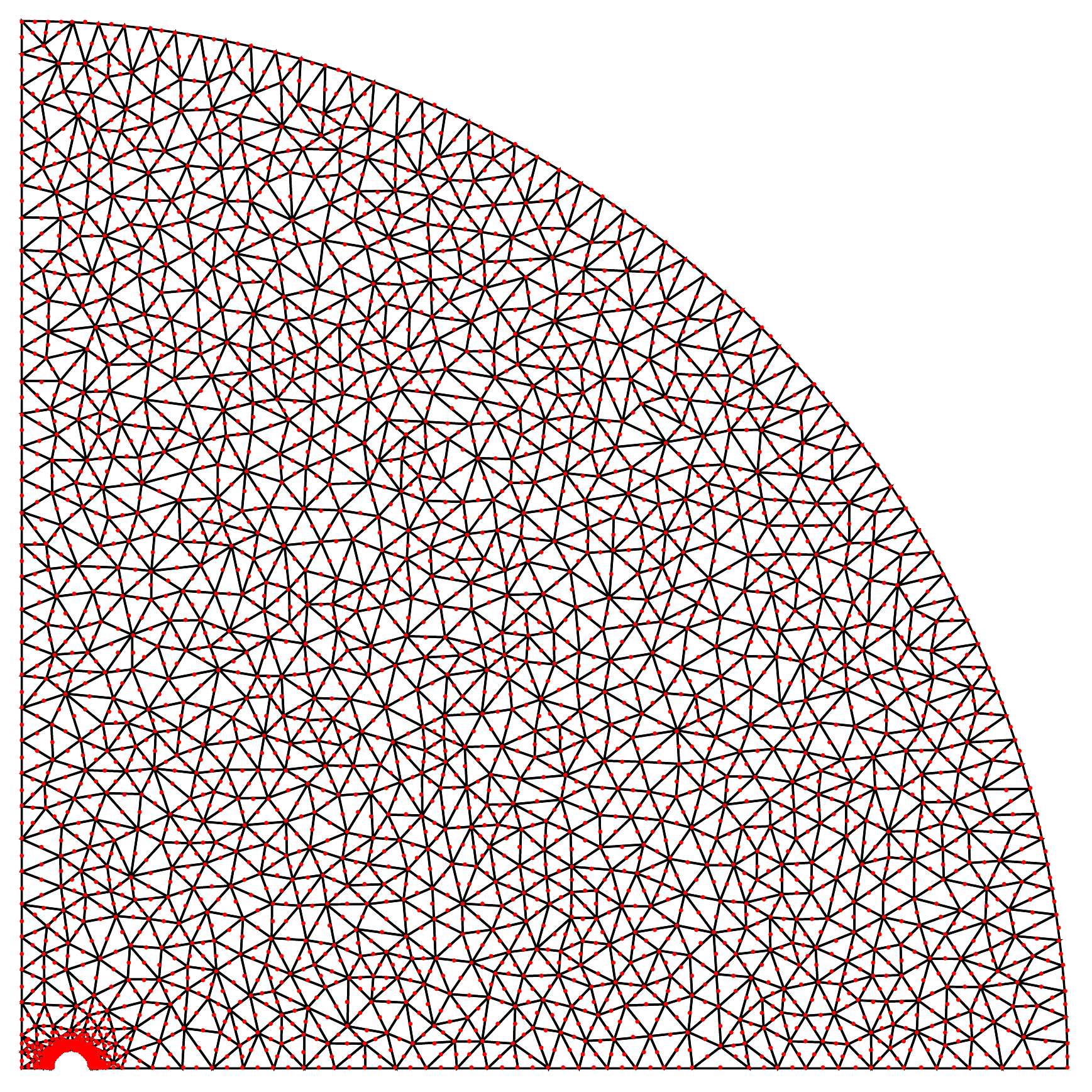}\qquad
	\includegraphics[width=.35\textwidth]{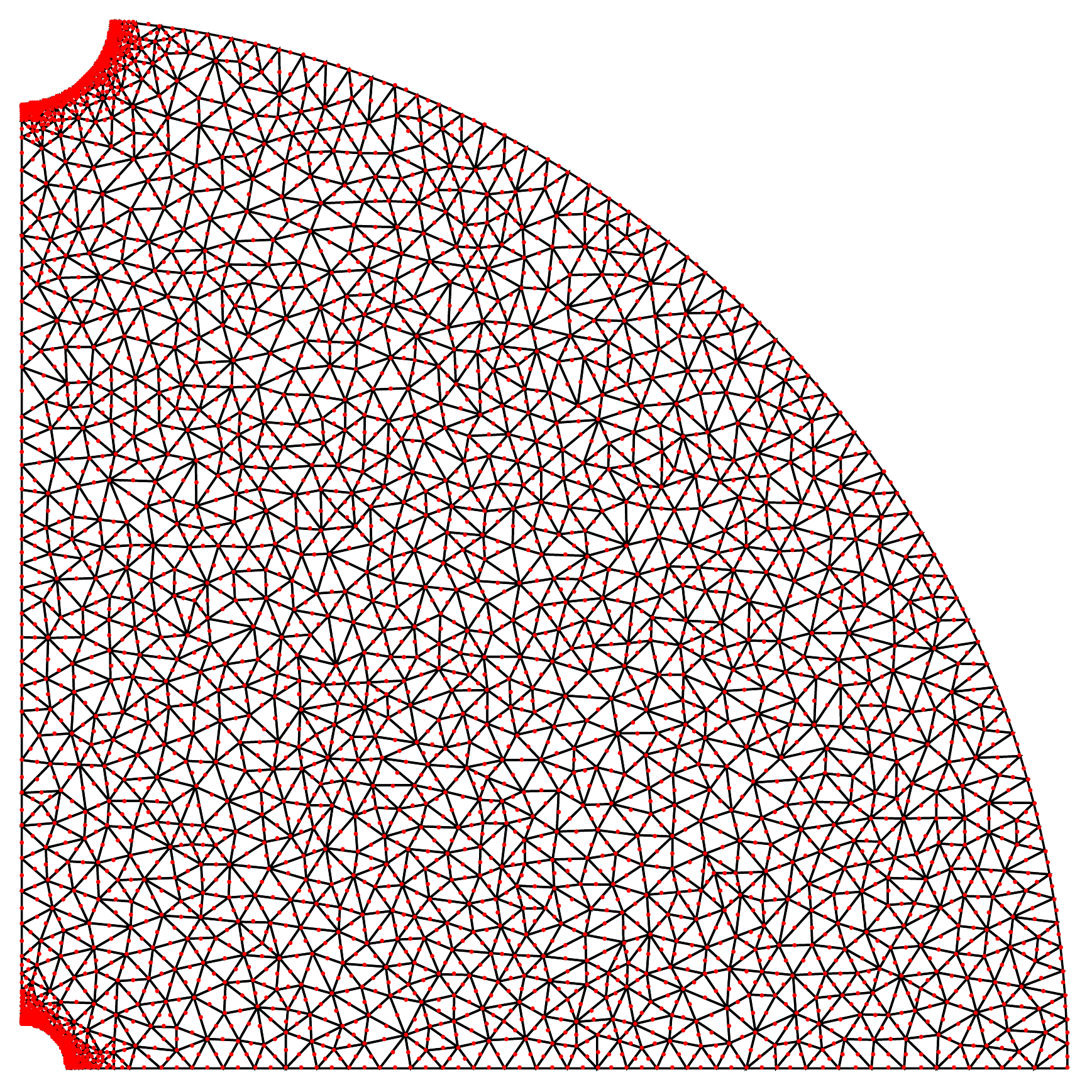}
\caption{Examples of the meshes used for the two types of Schottky group, in this case at the triple point where all phases meet.\label{mesh}}
\end{figure}

Following the finite element method, we decompose the domain into a mesh of triangular `elements'.\footnote{Common mesh generation algorithms include Chew's second algorithm and Ruppert's algorithm for generating (constrained) Delaunay triangulations.  We simply use internal mesh generation routines in \texttt{Mathematica 10}'s FEM package.}  The decomposition must be a valid triangulation in that triangles must meet edge to edge and vertex to vertex. On the boundaries, we allow the triangles to have curved edges.  On each vertex and the midpoint of each edge, place a node.  Each triangular element therefore has six nodes, and these nodes are shared between adjacent elements. Examples are shown in figure \ref{mesh}.

With a valid mesh, we can now define our Sobolev space.  For each of the nodes $i$, we can define a function $\psi_i$ that satisfies  $\psi_i(x_j,y_j)=\delta_{ij}$ for all nodes $j$ with coordinates $(x_j,y_j)$.  In two dimensions, a second-order polynomial is uniquely specified by its value on six points, and each element contains six nodes.  We can therefore extend $\psi_i$ to non-nodal points via polynomial interpolation within each element. This yields a set of piecewise continuous functions $\psi_i$ that are second-order polynomials in each element. The span of all $\psi_i$ together with the standard integral $L^2$ norm forms a Sobolev space which is our space of functions.  

Given the values of any $C^2$ function $\phi$ on the nodes $\phi_k=\phi(x_k,y_k)$, we can approximate $\phi$ using an element of the Sobolev space as $\phi(x,y)\approx \sum_{k}\phi_k \psi_k(x,y)$. By construction, this approximation agrees with the true function $\phi$ on the nodes. The approximation can be improved by making the mesh finer.  Because these are second-order polynomials, the approximation should improve roughly cubically with the number of elements, assuming an evenly distributed mesh.

Now let us describe how to solve the integral form of the linear equation above. We decompose the integrals as a sum of integrals on each element, and approximate all functions as functions on our Sobolev space.  Within each element, integrals and derivatives can be computed from second-order polynomials\footnote{In practice, integrals are computed using Gaussian quadrature, and differentiation matrices are derived by differentiating the $\psi_i(x,y)$ on each of the nodes.}.  If we take as a test function a basis function $\psi_i$, the integral equation reduces to an equation of the form $\sum_jA_{ij}\delta\phi_j=b_i$.  Over all basis functions, this defines a matrix $A_{ij}$ and a vector $b_i$, which gives a sparse linear algebraic system that can be solved using standard algorithms such as multifrontal methods. 

We note that the nodes that lie on the boundary of the domain $D$ contribute to the boundary terms in the integral equation. The fact that the interior nodes have no such contribution is equivalent to requiring that outward-pointing normal derivatives cancel between adjacent elements. This enforces the continuity of the first derivatives of the functions in the fine mesh limit.  (Continuity of the functions themselves was already guaranteed by the piecewise construction of the Sobolev space.) This is why the second-derivative terms were integrated by parts in the integral equation. 

\begin{figure}[htbp]
 \centering
	\includegraphics[width=.45\textwidth]{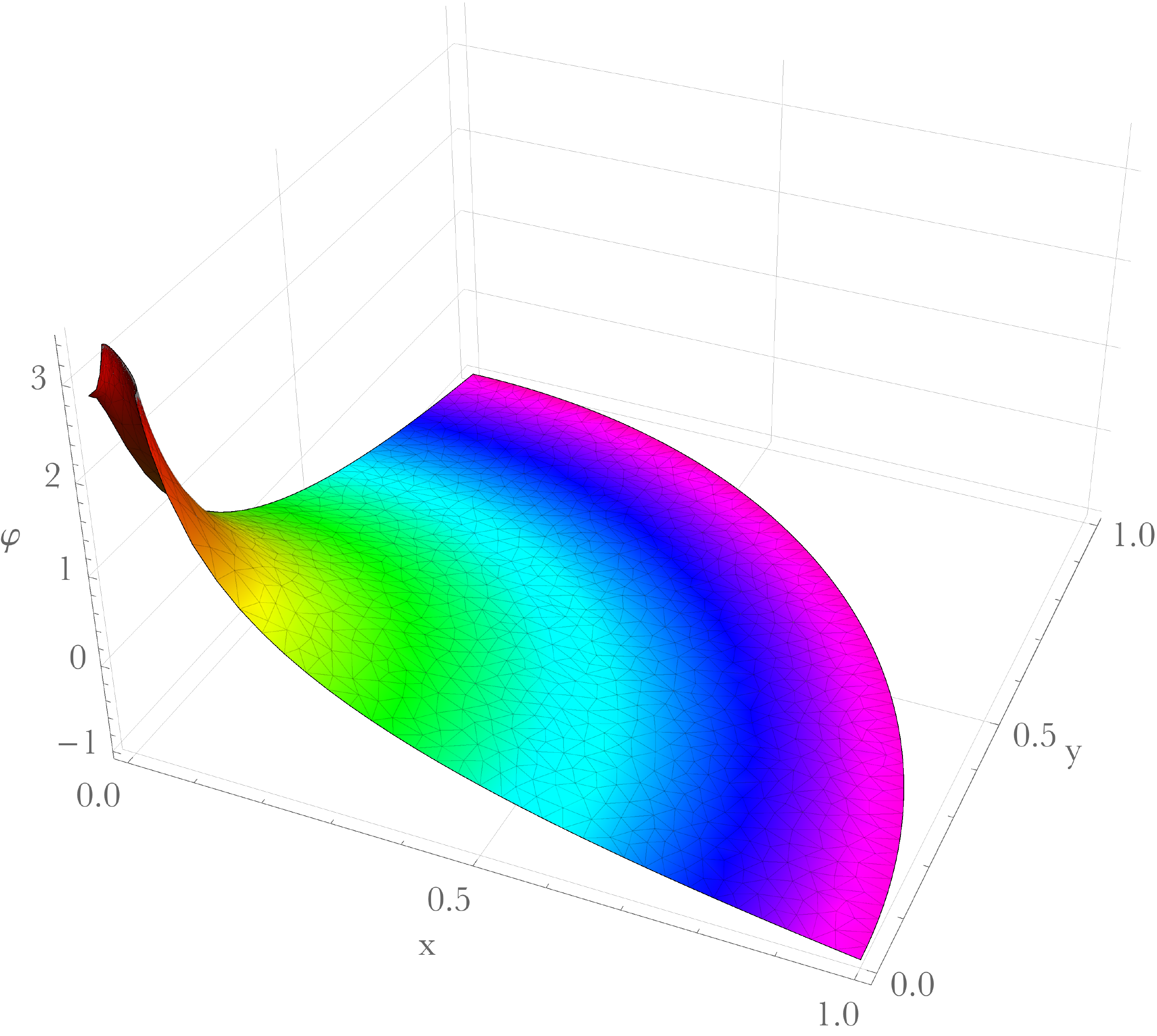}\qquad
	\includegraphics[width=.45\textwidth]{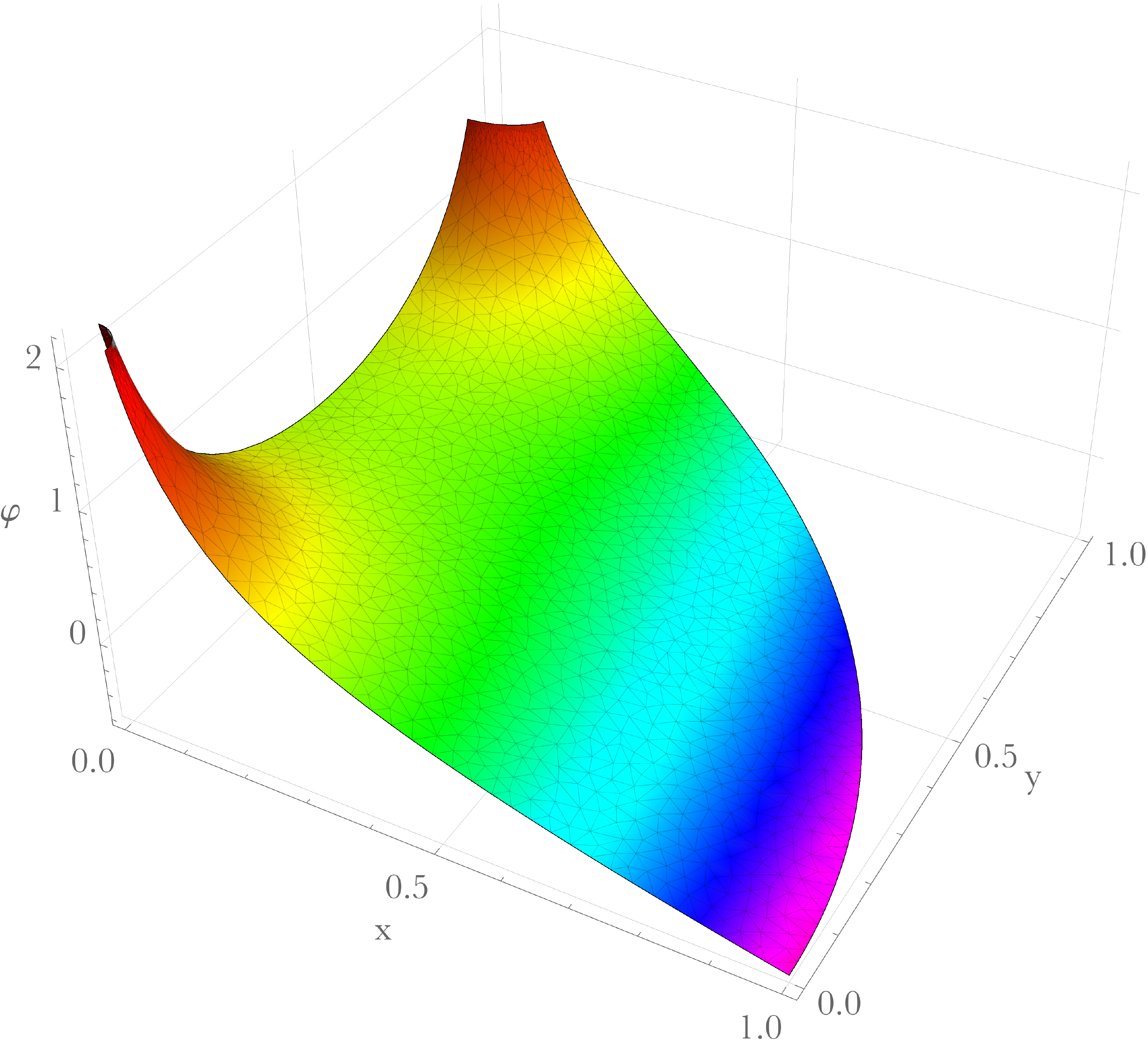}
\caption{Plots of the solutions of the Liouville equation for each of the phases, at the triple point where all phases meet.}
\end{figure}

\begin{figure}[htbp]
	\centering
	\includegraphics[width=.7\textwidth]{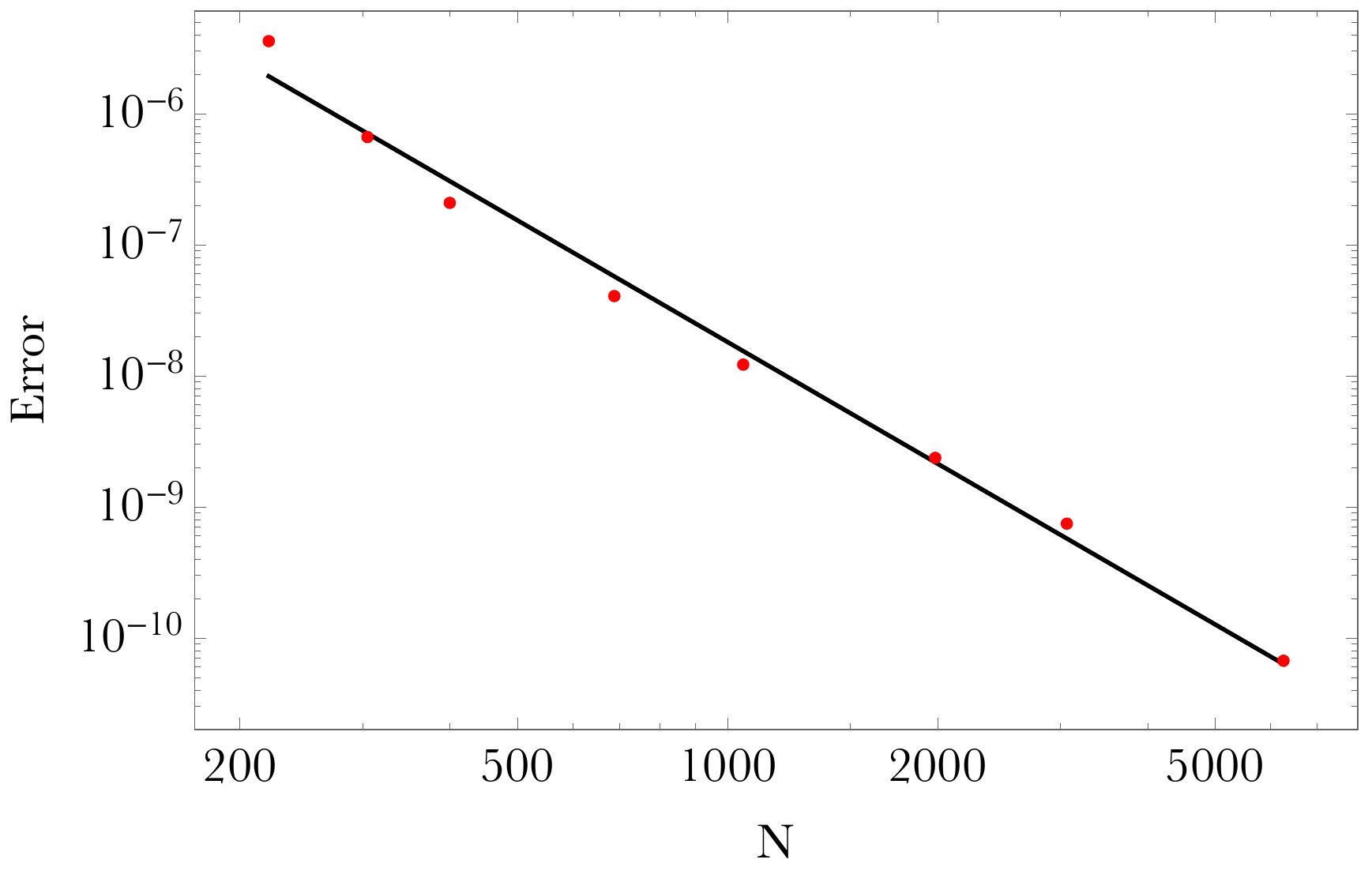}
	\caption{Log-Log plot of error in the Gauss-Bonnet theorem: $1-\frac{A}{4\pi}$, as a function of the number of elements $N$. The linear fit gives a power $N^{-3.08}$.\label{fig:convergence}}
\end{figure}

In Fig.~\ref{fig:convergence}, we show the convergence of this numerical method.  As a measure of error, we compute the area of the entire Riemann surface under the metric \eqref{smet}.  According to the Gauss-Bonnet theorem, this area must be $4\pi$ on a genus-2 surface with the constant curvature metric $R=-2$. The convergence is cubic in the number of elements, as predicted by the method. 

The solutions and the integration domain are parametrised by the various Schottky parameters described in appendix \ref{integrals}.  Horizon lengths can be obtained from these parameters via \label{eq:geoLengths}.  From each solution, we obtain the cycle lengths $\ell_a$ of the A-cycles and B-cycles by numerically integrating $\int_C e^\phi $ along the curves  described in figure \ref{SchottkyDomains}, which are geodesic from symmetry. This allows us to match the moduli of Schottky groups in different phases. We then compute the on-shell bulk action $I$ from the integrals in appendix \ref{integrals}. Since the bulk saddles corresponding to connected and disconnected phases are related by the additional $\mathbb Z_2$ symmetry, there are two families of solutions we must find in order to complete the phase diagram.  One family allows us to obtain both the connected wormhole phase and the fully disconnected phase.  Each solution in this family will have one value of the action, but two different set of lengths $\ell_a$, depending on which $\mathbb Z_2$ symmetry is viewed as a time reflection symmetry.  The second family corresponds to the partially connected phases with AdS and a BTZ black hole.  

\section{Results for genus two}
\label{res}

Consider first the subspace with cycle lengths $\ell\equiv\ell_1=\ell_2=\ell_3$, which has a $\ZZ_3$ symmetry exchanging the boundaries. There are two phases respecting the full symmetry group, corresponding to connected and disconnected spacetimes, and we also consider a partially connected phase breaking the $\ZZ_3$ symmetry.  We can directly restrict the parameters of our Schottky groups to this symmetry in the totally connected/disconnected families, though in the partially connected family it is not manifest which Schottky parameters correspond to equal cycle lengths, so we must scan the two-dimensional $\ell_1=\ell_2$ space of parameters and use interpolation to obtain actions and cycle lengths for the $\ell\equiv\ell_1=\ell_2=\ell_3$ family.

\begin{figure}[htbp]
	\centering
	\includegraphics[width=.7\textwidth]{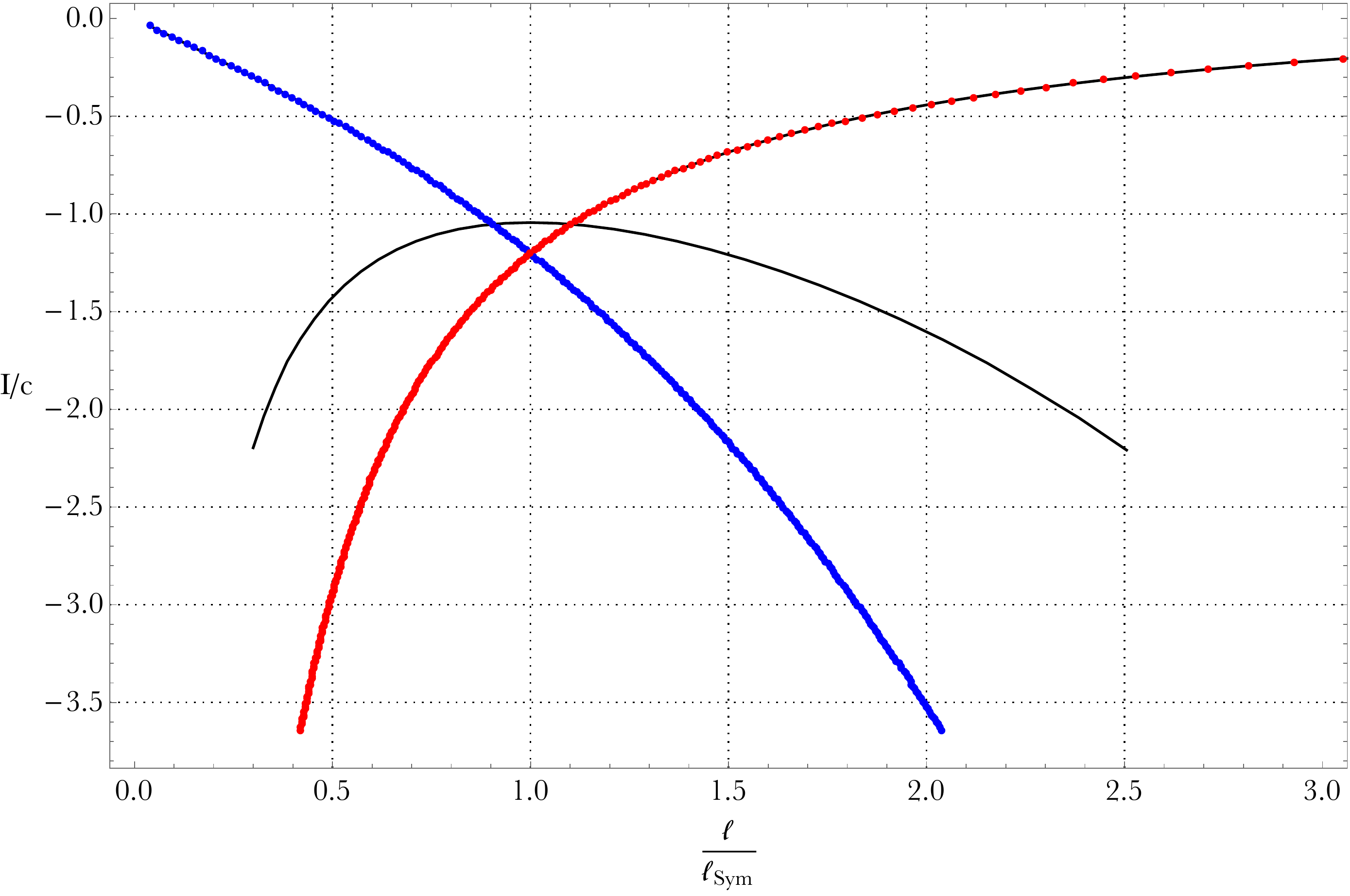}
	\label{fig:actions} 
	\caption{Bulk action as a function of cycle lengths $\ell\equiv\ell_1=\ell_2=\ell_3$, normalised to that of the most symmetric surface $\ell_{\mathrm{sym}}\equiv2\log(2+\sqrt 3)$. The phase with the lowest action is preferred.  At smaller lengths, the totally disconnected (three copies of AdS) phase is preferred (red in colour version), while for larger lengths, the connected wormhole is preferred (blue in colour version).  The phase transition between these occurs precisely at the length $\ell_{\mathrm{sym}}$. The AdS plus BTZ phase (black in colour version) is never dominant. All phases have a lower action than the non-handlebody phases ($I/c=0$.)}
\end{figure}

In Fig.~\ref{fig:actions}, we plot the bulk action as a function of $\ell$. We find that the AdS plus BTZ phase is never dominant.  At smaller lengths, the totally disconnected (three copies of AdS) phase is preferred, while for larger lengths, the connected wormhole is preferred.  The phase transition between these phases occurs for the maximally symmetric real Riemann surface where $\ell=\ell_{\mathrm{sym}}=2\log(2+\sqrt 3)$, which corresponds to horizon sizes $\lambda_{1,2,3}\approx 7.30$, slightly larger than the $2\pi$ horizon length corresponding to the Hawking-Page transition for the thermofield double.

Generalising to the two-dimensional family with $\ell_{12}\equiv\ell_1=\ell_2$, all three phases we consider are dominant for some region of moduli space.
\begin{figure}[htbp]
	\centering
	\includegraphics[width=.7\textwidth]{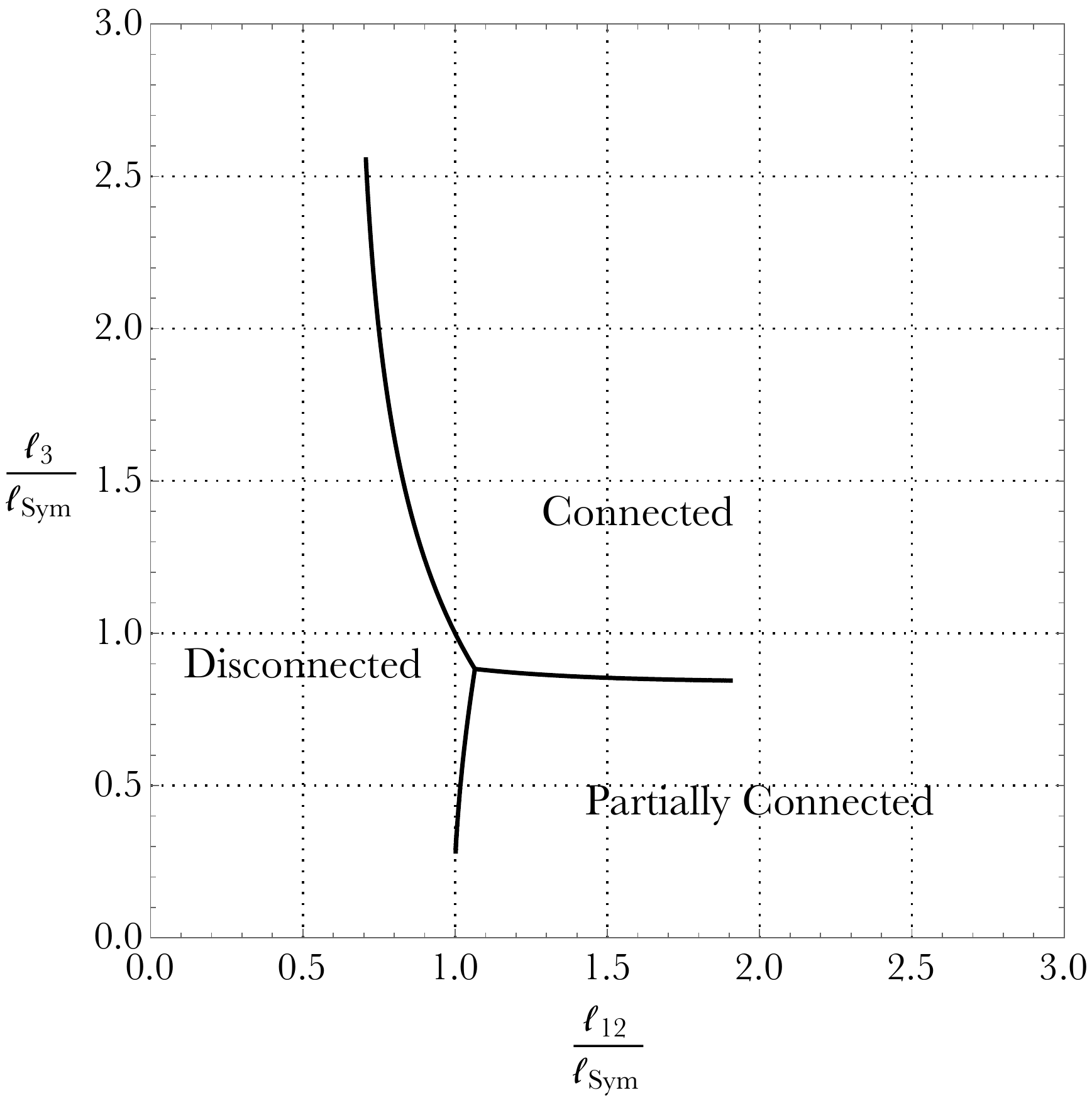}
	\caption{Phase diagram over cycle lengths $\ell_{12}\equiv\ell_1=\ell_2$ and $\ell_3$, normalised to that of the most symmetric surface $\ell_{\mathrm{sym}}\equiv2\log(2+\sqrt 3)$. The partially connected phase shown here connects the two $\ell_1=\ell_2$ cycles together.  The remaining partially connected phase is never dominant.\label{fig:phasediagram}}
\end{figure}
In Fig.~\ref{fig:phasediagram}, we show the phase diagram of solutions in this family, as a function of the cycle lengths of the Riemann surface.  The phase boundaries are found using interpolation. We see that the disconnected phase is preferred whenever $\ell_{12}$ is small, the partially connected phase (where boundaries with equal lengths are connected) is preferred whenever $\ell_{12}$ is large and $\ell_3$ is small, and the connected phase is preferred when all lengths are large. We also checked the handlebodies corresponding to a partially connected phase with the third boundary connected to one of the first two, which would break the symmetry swapping boundaries one and two (interpreted as time-reversal for the torus wormhole), finding they are never preferred.
We also find that all of these phases have an action below that of the non-handlebody solution (with $I=0$) throughout moduli space, though the subdominant solution appears to approach zero at the edges of moduli space.

The boundary between the connected and disconnected phases lies on the line of enhanced symmetry given by equation \eqref{eq:conDiscTransition}, where the A- and B-cycles have equal sizes, though connected and disconnected solutions spontaneously break this symmetry. The phase boundary ends at the tricritical point, where the partially connected phase takes over, and the handlebody respects the symmetry, acting by swapping the two generators (demanding $\alpha=\hat{\alpha}$ in the notation of the appendix).

\begin{figure}[htbp]
	\centering
	\includegraphics[width=.7\textwidth]{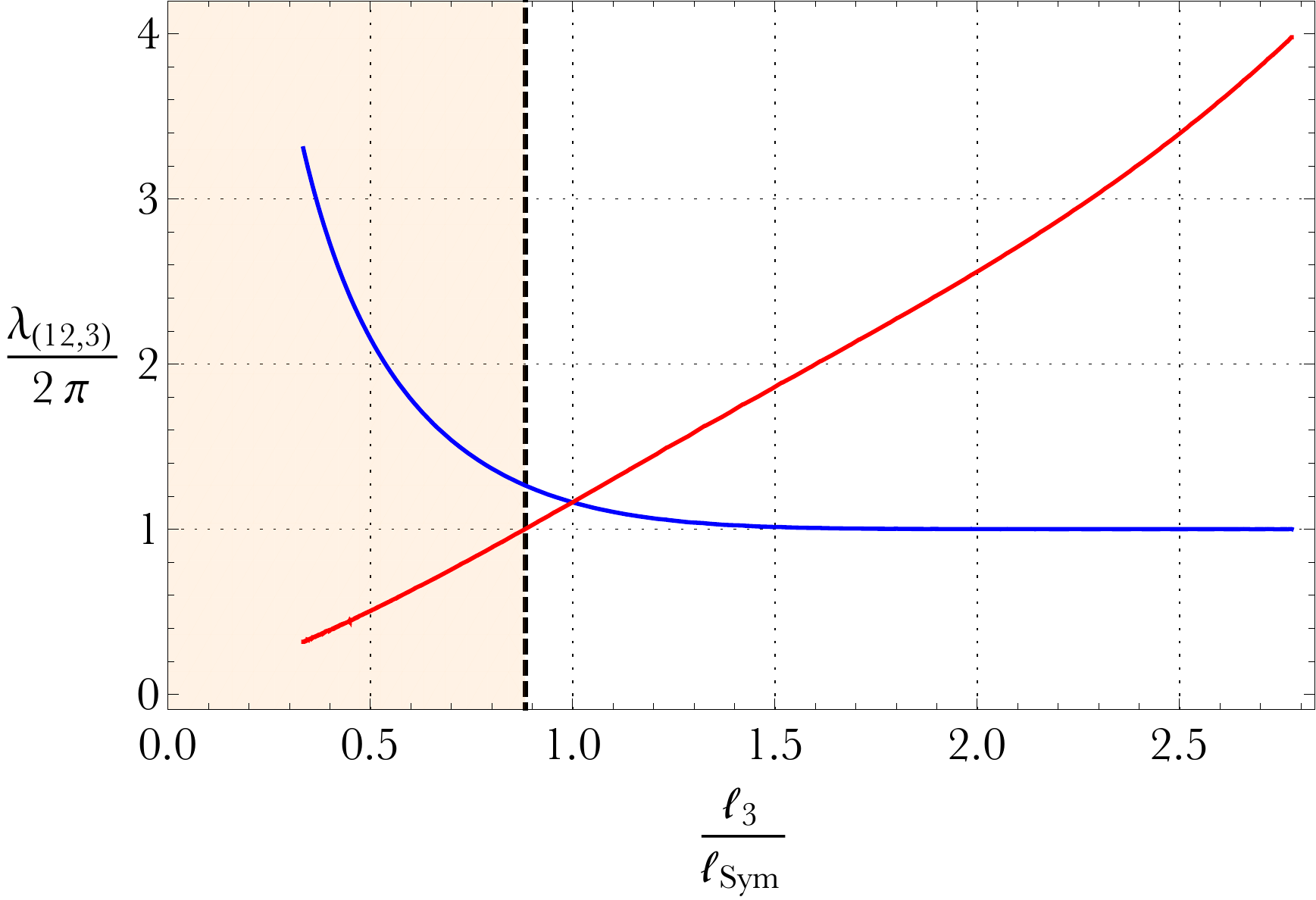}\vspace{1cm}
	\includegraphics[width=.41\textwidth]{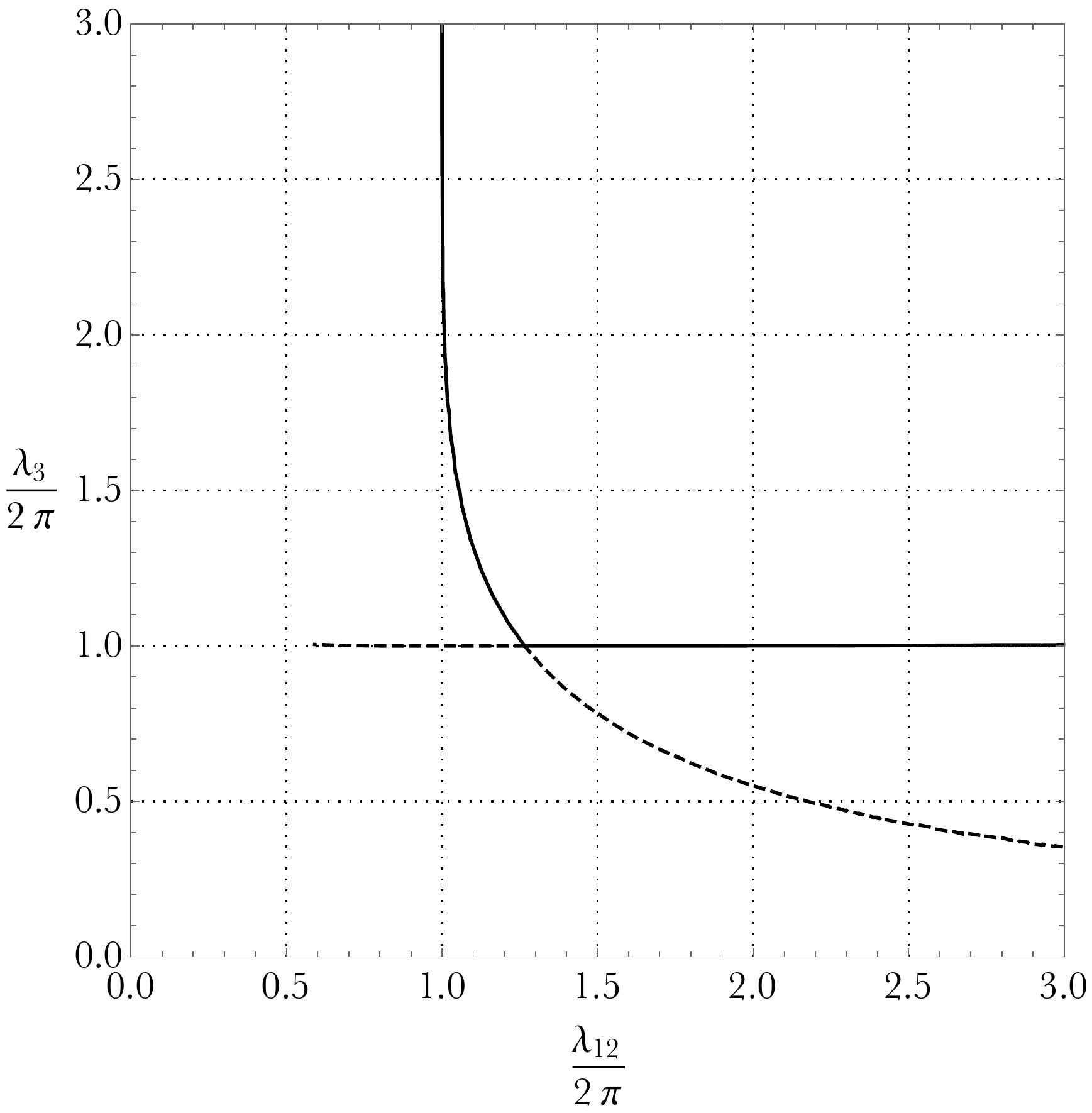}\qquad
	\includegraphics[width=.42\textwidth]{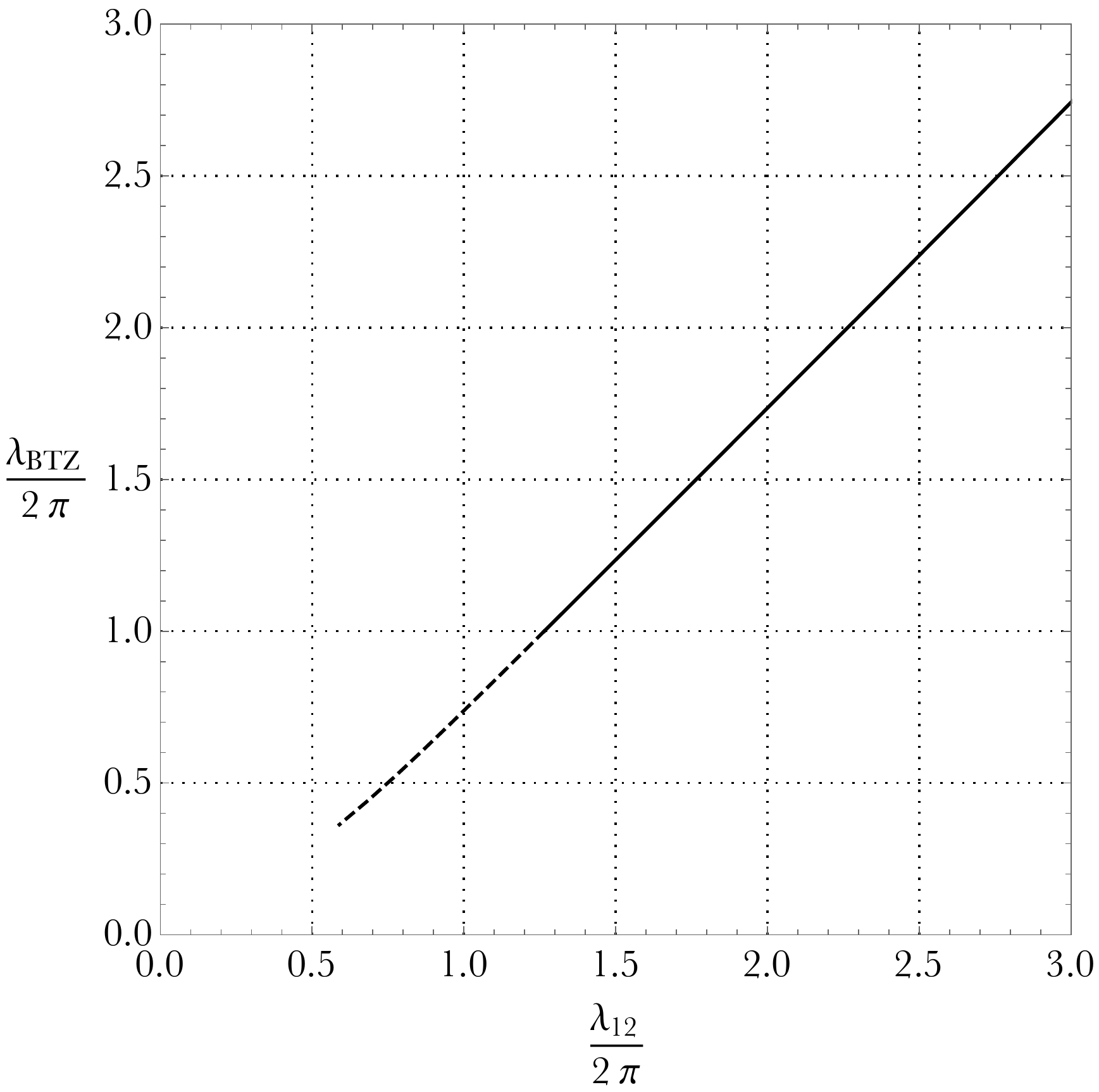}
	\caption{The horizon lengths in the connected phase at the boundaries where it meets the other phases. The upper plot shows $\lambda_{12}$ (blue) and $\lambda_3$ (red) as a function of the cycle length $\ell_3$, along the boundary with the disconnected phase; in the shaded part to the left of the dashed line, the partially connected phase is dominant. The lower left diagram shows the horizon lengths at which the connected phase becomes subdominant to the other phases, so the allowed horizon lengths in this phase are in the region in the top right, bounded by the solid lines. The lower right diagram shows the BTZ horizon length $\lambda_\text{BTZ}$ in the partially connected phase where it transitions to the connected phase with horizon lengths $\lambda_{12}$ (and $\lambda_3\approx 2\pi$, not shown); these have physical interpretations as the entropy of boundaries 1 and 2 either side of the transition. In the lower diagrams, dashed curves indicate where the two phases involved in the transition are both subdominant to the third phase.\label{fig:horizonPhases}}
\end{figure}

The corresponding horizon sizes for the connected phase at the phase boundaries, determined by the Schottky parameters, are shown in figure \ref{fig:horizonPhases}. In particular, along the phase transition between connected and disconnected phases, taking $\ell_3\to\infty$, the third horizon $\lambda_3$ in the connected phase is large but the lengths $\lambda_{12}$ of the two equal horizons approach $2\pi$ from above.  

At the phase boundary between the connected and partially connected phase (see lower left plot of figure \ref{fig:horizonPhases}), we find the striking result that the horizon length $\lambda_3$ of the connected phase is $\lambda_3\approx 6.283$, agreeing with $2\pi$ to high accuracy (to around one part in $10^4$ at the triple point), along the entire phase boundary.  A similar agreement with $2\pi$ appears in the partially connected phase as well, where the length of a geodesic circling in the Euclidean time (homotopic to the third B-cycle) agrees with $2\pi$ to a similar accuracy.  Swapping A- and B-cycles to reinterpret this as the disconnected/partially connected phase boundary, this geodesic becomes the horizon of the BTZ part of the partially connected geometry, so the transition from partially connected to disconnected appears for approximately the same BTZ horizon size as the Hawking-Page transition.

In terms of the torus wormhole, the interpretation of these results is that the black hole phase is dominant roughly when both the internal cycles of the torus have length of at least $2\pi$. In particular, the minimal horizon length, occurring for the square torus at the triple point, is at $\lambda\approx 22.3$, much greater than the Hawking-Page value.

On more careful investigation, this phase transition does not correspond to horizon sizes of \emph{exactly} $2\pi$, except perhaps in the limit $\ell_{12}\to\infty$ at the edge of moduli space. We checked this by setting the various $\lambda$'s in question to $2\pi$ and scanning the resulting one-parameter families of solutions.  We find that for sufficiently small $\ell_{12}$, these lines do not have matching cycle lengths, though the deviations quickly become very small as $\ell_{12}$ increases.  In particular, this implies that if the connected/partially connected phase boundary were extended much farther into the region where the disconnected phase is dominant, then the horizon lengths will show appreciable deviations from $2\pi$. The lack of matching is particularly obvious in the limit $\ell_{12}\to 0$, as one of the proposed families has $\ell_3\to\infty$ while for the other $\ell_3$ remains finite in the limit. We do not yet have an explanation for why the lengths approximate $2\pi$ so well for such a large range of parameters. 

\begin{figure}[htbp]
	\centering
	\includegraphics[width=.45\textwidth]{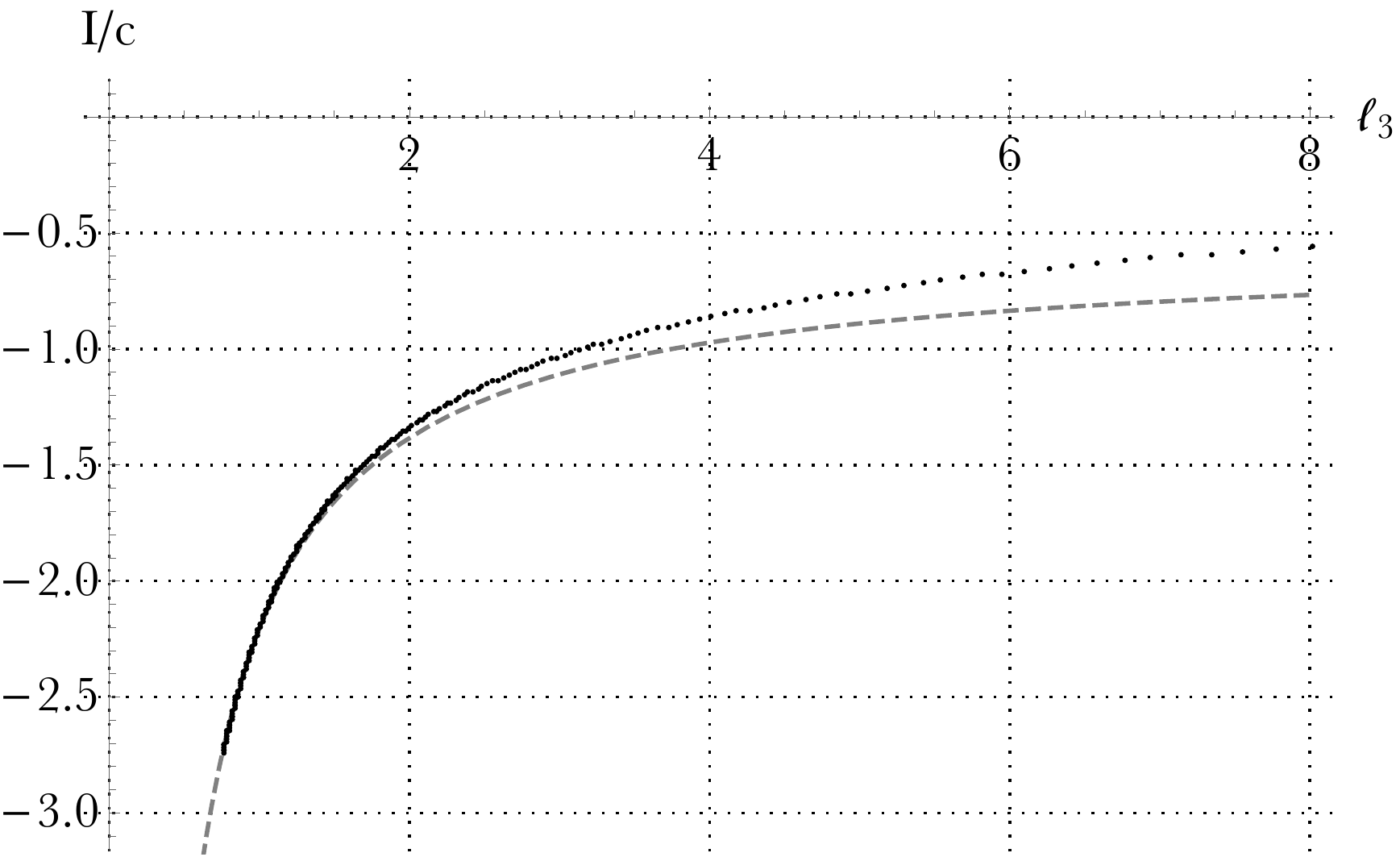}\quad
	\includegraphics[width=.45\textwidth]{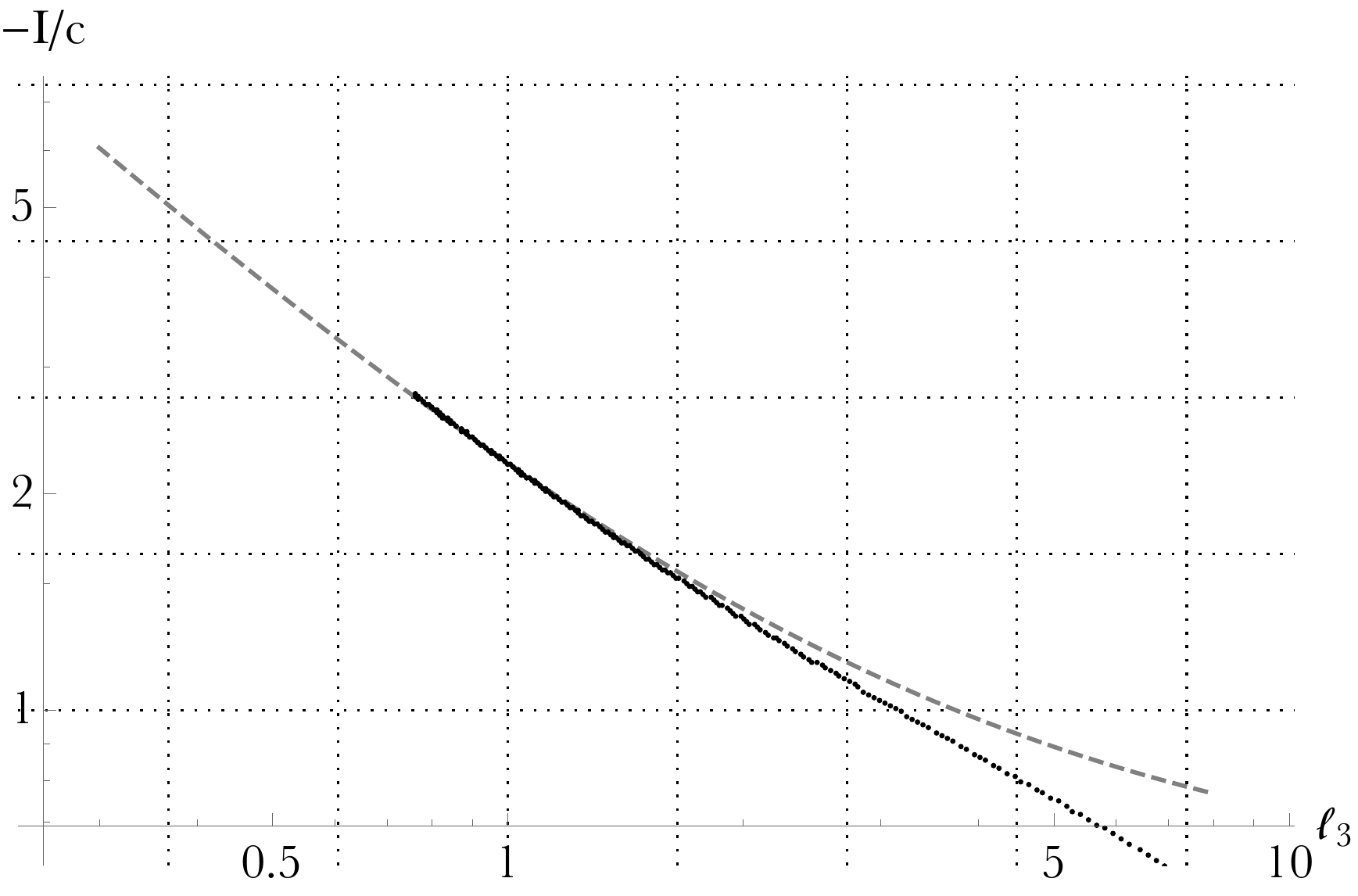}
	\caption{The action at the phase transition between partially connected and disconnected phases, compared with the analytic result $I/c=-\frac{\pi^2}{6\ell_3}$ in the $\ell_3\to0$ limit (dashed curve). A constant has been added to the analytic result, chosen to fit the data. The right plot shows the same data on a log-log scale.\label{pinchData}}
\end{figure}

Finally, we compared our numerical results with the pinching limits analytically obtained in section \ref{sec:pinching}. The numerics becomes difficult for small cycle lengths, due to a large separation of scales in the domain, so this check is not very precise, though we find good agreement. For example, taking the pinching limit $\ell_3\to0$ along the line separating disconnected and partially connected phases, a fit of the action to a function $-A \frac{\pi^2}{6}\ell_3^{-p}+C$ using 64 data points with $0.76\lesssim \ell_3 \lesssim 1$ gives parameters $p\approx 1.03, A\approx .99$. Data for this pinching limit is shown in figure \ref{pinchData}, plotted against $-\frac{\pi^2}{6\ell_3}+C$ with a one-parameter fit of the additive constant.

\section{Discussion}

A Riemann surface of genus $g \geq 1$ can be described by Schottky uniformisation as a region in the complex plane with identified boundaries. This Schottky uniformisation description can be extended to a quotient of $\mathbb H^3$, giving a handlebody solution in a Euclidean bulk, holographically dual to the field theory on the Riemann surface. A given Riemann surface has infinitely many Schottky uniformisations; from the bulk perspective these correspond to different handlebodies, characterised by the choice of boundary cycles which become contractible in the bulk. The geometry of least action dominates in the calculation of the partition function. 

In \cite{Krasnov:2000zq}, the calculation of the action of the bulk handlebodies was reduced to the calculation of the Takhtajan-Zograf action \cite{TZ} for a boundary Liouville field, describing the conformal map from the flat metric on the complex plane to the chosen metric on the Riemann surface. We have shown how, choosing a canonical constant negative curvature metric, this Liouville field may be computed numerically using finite element methods, and the Takhtajan-Zograf action evaluated. Apart from the importance for holographic studies, the solution for the Liouville field may be of some mathematical interest. 

We explicitly calculated the action for some bulk handlebody solutions in a two-dimensional subspace of the moduli space of genus two Riemann surfaces with three commuting reflection symmetries. Restricting to handlebodies preserving these symmetries allows us to consider finitely many solutions, three in this case. These surfaces have two distinct interpretations as preparing Lorentzian states on the circles of reflection symmetry: they give a state in three copies of the CFT, or a state in a single copy of the CFT. In the first interpretation, the Lorentzian analogues of the handlebodies we consider are a three-boundary wormhole, a totally disconnected space, and a two-boundary wormhole plus a disconnected space. In the second, they are two different versions of pure AdS and a single-exterior black hole with a torus behind the horizon. We also compared these solutions to a non-handlebody solution obtained as the quotient of a Euclidean wormhole. 

We compared the actions for these different solutions and determined the dominant solution as a function of the two moduli. We confirmed the accuracy of the numerical results by comparing to analytic expectations at special values of the moduli, and in pinching limits at the boundary of the moduli space. There is a line in the moduli space with a $\mathbb Z_3$ symmetry; along this line, we found that the $\mathbb Z_3$ symmetric saddles dominated, and there is a Hawking-Page like phase transition between the connected and disconnected phases. However, along another line in the moduli space with an enhanced $\mathbb Z_2$ symmetry, preserved by the partially connected phase but not the others, the symmetric phase does not always dominate. The full phase diagram is given in figure \ref{fig:phasediagram}. The handlebodies always dominate over the non-handlebody solution. 

Our results largely confirm previous expectations, but it is worth noting that symmetric solutions are not always dominant over those which spontaneously break the symmetry. This is similar to the square torus, the critical surface at the Hawking-Page phase transition, with a similar extra symmetry spontaneously broken by both dominant phases, although in that case there exists no handlebody that respects the full symmetry group of the boundary. It would be interesting to understand which symmetries are always preserved and which may be spontaneously broken. We should note that an important class of symmetries, which have interpretations as time-reversal ($\epsilon=0$ type real structures, in the language of section \ref{surf}), are found never to be broken in our examples, given the phases we have checked. Such a symmetry breaking would make semiclassical Lorentzian interpretation of the dual geometry more difficult, since there would be no natural $t=0$ slice on which to define the Hartle-Hawking state. Furthermore, if all such symmetries are preserved, it implies that there is no breaking of replica symmetry in calculations of R\'enyi entropies (at least for time-reversal invariant states and subsystems), since the dihedral replica symmetry group can be generated by such reflections.

We found that along the boundaries separating the partially connected phase from the others, certain horizon lengths approach $2\pi$ at the edges of moduli space, and this remains a very good approximation well into the centre of moduli space. It would be interesting to analytically understand these edges of moduli space, including the corrections to explain why the approximation remains good in a regime na\"ively so far from the limit. This would no doubt be useful for more general Riemann surfaces, and since the approximation is valid in such a wide parameter range there is more hope for good analytic understanding than one might have expected. A direct CFT calculation may also be tractable in such limits, giving an interpretion of the phases as dominance of the vacuum Virasoro block propagating in certain cycles.

An obvious avenue for future work is to extend to higher genus, where an increased number of boundaries allows for a richer multiparty entanglement structure. It is in fact possible to deduce results at higher genus for some special moduli already without further work via a trick, whereby we may extend a pair of pants decomposition of a surface into the bulk, cutting certain handles of a handlebody, and getting the action for each solid pair of pants from the genus 2 results. This is usually not possible because the cutoff surface chosen (or equivalently the Liouville field) depends on the surface globally, and this affects the action of each component. However, if the cycles we cut along are fixed by a reflection symmetry, the cutoff in each pair of pants component is identical to the cutoff for half the genus two surface, so the action can be found. In terms of the Schottky domain, this is manifested by the domain $D$ being divided into pairs of pants bounded by geodesics (at known locations fixed by symmetries), which allows solution of the Liouville equation without reference to the identifications used. As a simple example of the sort of result that  can be deduced from this idea, the four boundary wormhole with equally sized boundaries arranged to have the symmetries of a square (boundaries lying in a plane, one at each vertex of the square, roughly speaking) has a transition from connected to disconnected phases at horizon sizes $\lambda\approx 7.62$.

 One application of the results is to see what the phase structure implies for the possible structure of entanglement in these states. For example, a phase transition may prevent the moduli from ever exhibiting intrinsically $n$-party entanglement in the sense of \cite{Balasubramanian:2014hda}, or, more generally, may restrict the entropy cone of \cite{Bao:2015bfa}. The reason for this is that choosing horizon sizes large enough to be above the phase transition may imply that every hyperbolic surface has another internal cycle short enough to enter a Ryu-Takayanagi calculation. For example, taking the four boundary wormhole with the symmetries of a square as in the previous paragraph, the internal moduli are relevant for horizon sizes $\lambda \gtrsim 2.12$, far below the phase transition point deduced above.
 
 It would also be interesting to check whether geodesics that leave the constant time slice can ever be short enough to dominate the entanglement entropy, as discussed in \cite{Maxfield:2014kra}, where for the torus this possibility was shown to be prevented only by the phase structure.
 
\paragraph{Acknowledgements}
It is a pleasure to thank Alex Maloney for helpful discussions. We would like to thank the Centro de Ciencias de Benasque Pedro Pascual, where this project was initiated, for their hospitality.
HM was supported by an STFC STEP award, and receives support from the Simons Foundation.
 SFR is supported by STFC under grant number ST/L000407/1.  B.W. is supported by European Research Council grant no. ERC-2011-StG 279363-HiDGR.

\appendix

\section{Action integrals}\label{integrals}

\subsection{Takhtajan-Zograf action}

We consider a handlebody geometry obtained from the quotient of $H^3$ by the Schottky group $G$ generated by M\"obius maps $L_1,\ldots, L_g$. On the boundary Riemann sphere of $H^3$, pick a fundamental region $D$ for $G$, bounded by $2g$ curves $\partial D = C_1 + C_1' +\cdots + C_g + C_g' $, with $C_k' = -L_k(C_k)$. The notation here means that the curves inherit their orientation as the boundary of $D$, which means clockwise in the plane for curves bounding $D$ from the inside. The reverse orientation of a curve is indicated by a sign.  

If one of the curves bounds $D$ from the outside on the $w$ plane, so the fundamental domain is bounded, we can write the Takhtajan-Zograf action as
\begin{align}
	I_{TZ}[\phi] =& \int_D \frac{i}{2} \dd w \wedge \dd\bar{w} \;  \left(4\partial \phi \bar{\partial} \phi +e^{2\phi}\right) \\
	  &+\sum_k \int_{C_k} \left(2\phi -\frac{1}{2}\log |L'_k|^2 -\log|c_k|^2\right) \frac{i}{2}\left(\frac{L''_k}{L'_k} d w - \frac{\bar{L}''_k}{\bar{L}'_k} d\bar{w} \right) \nonumber
\end{align}
where $c_k$ denotes the bottom left component of the $SL(2,\RR)$ matrix for $L_k$. The sum runs over all $k$ such that $w_\infty\neq\infty$ (or $c_k\neq 0$), so if one transformation is a scaling, rotation and translation ($w\mapsto q^2w+w_0$), it does not contribute a boundary term.

We will find it convenient to write this in a more geometric way. For each generator $L_k$ of $G$, let $w_\infty^{(k)}$ be the point $-\frac{d_k}{c_k}$ mapped to infinity, and $\theta_\infty^{(k)}$ the angle measured from this point. Then
\begin{align}\label{TZgeometric}
	I_{TZ}[\phi] =& \int_D \dd^2w \;  \left((\nabla\phi)^2 +e^{2\phi}\right) \\
	  &+\sum_k \int_{C_k} \left(4\phi +4\log |w-w_\infty^{(k)}|\right)\dd\theta_\infty^{(k)} \nonumber
\end{align}
where again the sum runs over all $k$ such that $w_\infty\neq\infty$.

When $C_k$ is a circle, the final boundary integral independent of $\phi$ can be explicitly evaluated in terms of the geometry. We will do the case where $C_k$ bounds $D$ from the inside, so that $w_\infty^{(k)}$ lies inside $C_k$. On $C_k$, $|w-w_\infty^{(k)}|$ is determined from the cosine rule, as
\begin{equation}
	|w-w_\infty^{(k)}|^2-2D_k\cos\theta_\infty^{(k)}|w-w_\infty^{(k)}|+D_k^2-R_k^2=0\, ,
\end{equation}
where $R_k$ is the radius of $C_k$, and $D_k$ the distance between its centre and $w_\infty^{(k)}$. But this equation has a second solution: it corresponds to the point with $\theta_\infty^{(k)}$ shifted by $\pi$, with a sign. The product of $|w-w_\infty^{(k)}|$ for these two related points is therefore $R_k^2-D_k^2$, read off from the constant term of the quadratic. So if we add the integrands coming from $\theta_\infty^{(k)}$ related by a shift by $\pi$, we get just this constant, and then the integral straightforwardly gives
\begin{equation}\label{intTerm}
	\int_{C_k} 4\log |w-w_\infty^{(k)}|\;d\theta_\infty^{(k)} = -4\pi \log(R_k^2-D_k^2)
\end{equation}
with the sign coming from the clockwise orientation. We will be able to evaluate most other such integrals we encounter similarly.

We may also choose $D$ to be unbounded, including the point at infinity, but then the bulk integral diverges logarithmically. We therefore must integrate over a compact region, bounded by some regulating curve $C_\infty$ (for example, a circle of large radius), add the boundary term
\begin{equation}\label{unboundedD}
	4\int_{C_\infty}\left(\phi +\log |w|\right)\dd\theta\,
\end{equation}
and take the limit where $C_\infty$ retreats to infinity. In this limit, the boundary term is independent of the choice of origin.

\subsection{Using symmetries}

In the case where the Schottky group has some symmetry respected by the choice of metric, the action can be obtained from a `symmetry reduced' domain $\tilde{D}$, a fundamental region for the extension of $G$ by its automorphisms. The action is not the integral of an intrinsic, local quantity on the surface, and $\phi$ transforms nontrivially under the symmetries, so it takes some work to identify the expression in the symmetry-reduced domain.

The only symmetries we will use will be reflections in some line, and inversions in some circle. For the former, it is clear that we can just do all integrals over one half of the plane, and double the result, since there will be equal contributions from each half, but this is not the case for inversions.

So consider inversion in a circle of radius $R_i$, and use polar coordinates $(r_i,\theta_i)$ centred on the circle of inversion, so the symmetry acts as $r_i\mapsto R_i^2/r_i$. The equivariance property of $\phi$, ensuring that the metric on the surface respects the symmetry, is
\begin{equation}
	\phi(R_i^2/r_i,\theta_i)=\phi(r_i,\theta_i)+\log\left(\frac{r_i^2}{R_i^2}\right)
\end{equation}
with derivative
\begin{equation}
	\partial_{r_i} \phi(R_i^2/r_i,\theta_i)= -\frac{r_i^2}{R_i^2}\left(\partial_{r_i}\phi(r_i,\theta_i)+\frac{2}{r_i}\right)
\end{equation}
and in particular, evaluating this at $r_i=R_i$ tells us that the radial derivative of $\phi$ on the circle of inversion itself is $-1/R_i$.

The integral of the `kinetic term' over $D$ reduces to the part $\tilde{D}$, lying on one side of the inversion circle, as
\begin{align}\label{symBulk}
	\int_D \dd^2w (\nabla\phi)^2 &= 2\int_{\tilde{D}} \left[(\nabla\phi)^2 +\frac{2}{r_i}\partial_{r_i}\phi +\frac{2}{r_i^2}\right] \nonumber\\
	&= 2\int_{\tilde{D}}\dd^2w(\nabla\phi)^2 + 4\int_{\partial\tilde{D}}\phi\; \dd\theta_i + 4\int_{\partial\tilde{D}} \log r_i \;\dd\theta_i
\end{align}
where the extra boundary terms come from the nontrivial transformation of the radial derivative of $\phi$.

Now if $D$ is unbounded, and we use the inversion to make the reduced domain $\tilde{D}$ bounded, the boundary term \eqref{unboundedD} from the large circle of radius $R_c$ implementing the cutoff can be mapped to a small circle of radius $R_i^2/R_c$, giving
\begin{equation}
	4\int \left(\phi(r=R_i^2/R_c,\theta_i) - \log\left(\frac{R_c^2}{R_i^2}\right) + \log R_c\right)\dd\theta_i
\end{equation}
which precisely cancels the same boundary term from transforming the bulk integral, due to the opposite orientations. The upshot is that we need not include a cutoff in the reduced domain, or any additional terms, simply integrating over $\tilde{D}$ including the origin.

It will also be useful for us to integrate the `kinetic' term by parts, since this will eliminate the need to take any derivatives of the numerical solution for $\phi$.
\begin{equation}
	\int_{\tilde{D}} \dd^2 w\; (\nabla\phi)^2 = - \int_{\tilde{D}} \dd^2 w\; \phi \nabla^2\phi + \int_{\partial\tilde{D}}\dd s\; \phi \nabla_n \phi
\end{equation}
In the first term, when $\phi$ solves the Liouville equation we can use it to replace the Laplacian. The boundary term requires the normal derivative of $\phi$, but we will be able to obtain it analytically as $\partial\tilde{D}$ will invariably be made up of arcs of circles in which there is an inversion symmetry. This is not necessarily the original inversion symmetry, but may be a new inversion resulting from conjugation by some element of $G$. We know the radial derivative of $\phi$ on such a circle, $\nabla_n \phi=\pm \partial_{r_i} \phi = \mp 1/R_i$, with the sign depending on whether $\tilde{D}$ lies inside or outside the relevant circle, and the length element is $\dd s = \pm R_i\;d\theta_i$. This allows us to replace each component of the integral with
\begin{equation}\label{byParts}
	\int \dd s\; \phi \nabla_n \phi = -\int\phi\;\dd\theta_i 
\end{equation}

We also have to consider the transformation of any boundary terms that lie outside $\tilde{D}$. We now move on to describe the two different families of Schottky groups we are considering, where we will work out the boundary transformations in each case.

\subsection{First domain}

This section collects all the relevant information for the domain corresponding to the connected and completely disconnected phases of the three boundary wormhole, and the pure $AdS$ phases of the torus wormhole.

In the symmetric two-parameter family we consider, there are three commuting anticonformal involution symmetries, which can be thought of as acting on the Riemann sphere by reflection in the three coordinate planes. The curves $C_1,C_2,C_1',C_2'$ can then be taken as equally sized circles with centres on the equator, at longitudes $\pm\alpha$ for $C_1,C_1'$ and $\pi\pm\alpha$ for $C_2,C_2'$ for some angle $\alpha\in(0,\pi/2)$, so the Schottky group identifies neighbouring circles. The reflection in the equatorial plane, corresponding to the time-reflection for the three boundary wormhole in the connected phase, maps each circle to itself, but flipped. The reflection swapping $C_1 \leftrightarrow C_1'$ and $C_2 \leftrightarrow C_2'$ corresponds to time reflection when the bulk is reinterpreted as the three boundary wormhole in the disconnected phase. The final reflection, being the time-reversal for the torus wormhole, swaps $C_1 \leftrightarrow C_2$ and $C_1' \leftrightarrow C_2'$. Relaxing the assumption of this final symmetry would allow us to describe the full three-parameter space of three-boundary wormholes by having different radii for the pairs $C_1,C_1'$ and $C_2,C_2'$. We stereographically project this sphere to the $w$-plane in such a way that the equator maps to the real axis, and the symmetries act as reflection in the axes and inversion in the unit circle, which we denote by $C$. The whole domain $D$ is unbounded, but we use the inversion to consider $\tilde{D}$ as the part within the unit circle.

We can write the generators as
\begin{equation}
	L_1(w) = \frac{1}{w^{(1)}_0} \frac{w-w^{(1)}_0}{w-w^{(1)}_\infty}\, ,\quad L_2(w) = \frac{1}{w^{(2)}_0} \frac{w-w^{(2)}_0}{w-w^{(2)}_\infty}
	\end{equation}
where $w^{(k)}_0,w^{(k)}_\infty$ are real, in order to be symmetric under complex conjugation. There is a single parameter redundancy here, since we can act with an $SL(2,\RR)$ transformation that fixes $\pm 1$ while retaining this form for the generators. To impose the full symmetry including reflection in the $x$-axis, swapping the two generators, we further require $w^{(2)}_0=-w^{(1)}_0$ and $w^{(2)}_\infty=-w^{(1)}_\infty$. We will look only at the first generator here, simplifying notation by writing $w_0=w^{(1)}_0>0$, $w_\infty=w^{(1)}_\infty>0$, with the second generator being treated identically.

There is a unique choice of curves $C_1,C_1'$ respecting the symmetries, with $C_1$ a circle centred at $w_0$, with radius $R=\sqrt{w_0(w_0-w_\infty)}$. In terms of the variables $(w_0,R)$, we can write the $SL(2,\CC)$ matrix generator
\begin{equation}
	g_1=\frac{1}{R} \begin{pmatrix} 1 & -w_0 \\ w_0 & R^2-w_0^2\end{pmatrix}
\end{equation}
and the range of parameters allowed is $0<R<w_0<1-R$ (or something more general if we relax the $x\mapsto -x$ symmetry). The $SL(2,\RR)$ matrix representations are useful for relating this parameterisation to the horizon geometries, using \eqref{eq:geoLengths}; the lengths $\lambda_1,\lambda_2,\lambda_3$ of the three horizons come from the generators $g_1$, $g_2$, and $g_3=-g_1^{-1}\cdot g_2$.  A useful fact is that on $C_1$, we have $\theta_0=\theta+\theta_\infty$, where $\theta_0$, $\theta$ and $\theta_\infty$ are respectively the angles measured from $w_0$, $0$ and $w_\infty$.

We do not have much work to do here with boundary integrals, since $C_k$ is a boundary of $\tilde{D}$. Using the action \eqref{TZgeometric}, the symmetry reduction \eqref{symBulk}, and the analytic integral \eqref{intTerm} inserting the distance between $w_\infty$ and $w_0$ as $D_1=R^2/w_0$, we have
\begin{align}
	S_{TZ} &= 2\int_{\tilde{D}}\dd^2w\left[(\nabla\phi)^2+e^{2\phi}\right] + 4\int_{\partial\tilde{D}}\phi \; \dd\theta + \int_{\partial\tilde{D}} 4\log |w| \;\dd\theta \nonumber \\
	&\qquad+ \int_{C_1}4\phi \; \dd\theta_\infty -4\pi\log\left(R^2-\frac{R^4}{w_0^2}\right) + (\text{terms from $C_2$}).
\end{align}
We further simplify by integrating the first term by parts, and use \eqref{byParts} for the resulting boundary terms. The boundary $\partial\tilde{D}$ consists of three cycles: the unit circle $C$ (which inherits anticlockwise orientation as the boundary of $\tilde{D}$), and $C_1$ and $C_2$. The integrals including $\phi$ on $C_1$ and $C_2$ get three contributions, from the original boundary term, the additional term from the symmetry, and the integration by parts, which combine very simply after using $\theta+\theta_\infty=\theta_0$. The terms without $\phi$ dependence can be explicitly evaluated. The final result, not necessarily imposing symmetry under reflection in the $x$-axis, reinstating independent radii $R_1,R_2$ and centres $w_0^{(1)},w_0^{(2)}$ for $C_1$ and $C_2$, is
\begin{align}
	S_{TZ} &= 2\int_{\tilde{D}}\dd^2w\left[-\phi\nabla^2\phi+e^{2\phi}\right] +2
	\int_C \phi \; \dd\theta \nonumber \\
	&\qquad +2\int_{C_1}\phi \; \dd\theta_0^{(1)} +2	\int_{C_2}\phi \; \dd\theta_0^{(2)} -8\pi\log\left(R_1 R_2\right).
\end{align}

To match moduli, we measure the lengths $\ell$ of the three A-cycles and three B-cycles on the constant curvature metric, via $\ell=\int e^\phi$ evaluated numerically on the appropriate curve. The A-cycles are the curves fixed by complex conjugation: the segment of the real axis between $C_1$ and $C_1'$, the similar part between $C_2$ and $C_2'$, and finally the union of the remaining intervals of the real axis. The B-cycles are made up of the points fixed by the inversion symmetry (time-reversal in the interpretation as the disconnected phase), which are the circles $C_1$ and $C_2$, and the unit circle $C$.

\subsection{Second domain}

In this section, we repeat the above analysis for the second sort of domain we use, corresponding to the partially connected phase of the three boundary wormhole, and the black hole phase of the torus wormhole.

Once again, in the symmetric family we consider the three commuting anticonformal involution symmetries, act on the Riemann sphere by reflection in the coordinate planes. The curves $C_1,C_2,C_1',C_2'$ in this case are taken be taken as circles with centres on the equator, where the $x$-axis and $y$-axis intersect the sphere. Take $C_1$ and $C_1'$ to have equal radii and lie on opposite sides of the circle, and similarly for $C_2,C_2'$. The reflection in the equatorial plane now corresponds to time-reversal for the torus wormhole, and the other reflections each have interpretations as time-reversals for the partially connected phases of three boundary wormholes.

After projecting to the $w$-plane, the symmetries are again represented by reflections in the coordinate axes and inversion in the unit circle. We may choose the generators, in a convenient parameterisation, as
\begin{equation}
	L_1(w) = \tan^2\left(\frac{\hat{\alpha}}{2}\right) w\, ,\quad L_2(w) =\frac{w-\cos\alpha}{1-\cos\alpha\,w}
\end{equation}
with $\alpha,\hat{\alpha}\in(0,\pi/2)$, and $\tan\alpha\tan\hat{\alpha}<1$. The points mapped to zero and infinity by $L_2$ are $w_0=\cos\alpha$ and $w_\infty=\sec\alpha$. The fundamental domain $D$ is bounded by circles $C_1$ and $C_1'$, centred at the origin with radii $\tan\left(\frac{\hat{\alpha}}{2}\right)$ and $\cot\left(\frac{\hat{\alpha}}{2}\right)$ respectively, and $C_2,C_2'$ centred at $\pm\sec\alpha$, both with radius $\tan\alpha$. The circle $C_2$ meets the unit circle orthogonally at $e^{\pm i \alpha}$. Again denoting the angles measured from $0,w_0,w_\infty$ by $\theta,\theta_0,\theta_\infty$, we have $\theta_\infty = \theta+\theta_0$. Reducing the domain using the inversion means we need only consider the region $\tilde{D}$ inside the unit circle, bounded by $C_1$, the arcs $\tilde{C}_2$ and $\tilde{C}'_2$ of $C_2,C_2'$ lying within the unit circle, and the two parts of the unit circle between $C_2$ and $C_2'$, which we denote by $C$, oriented anticlockwise as inherited from $\tilde{D}$.


The $SL(2,\CC)$ generators of the Schottky group are then
\begin{equation}
	g_1=\begin{pmatrix} \tan\left(\frac{\hat{\alpha}}{2}\right) & 0 \\ 0 & \cot\left(\frac{\hat{\alpha}}{2}\right)\end{pmatrix}
, \quad g_2=\frac{1}{\sin\alpha} \begin{pmatrix} 1 & -\cos\alpha \\ -\cos\alpha & 1\end{pmatrix}
\end{equation}
and are again useful for relating $\alpha,\hat{\alpha}$ to the parameters of the bulk geometry by equation \eqref{eq:geoLengths}. Interpreted as the partially disconnected phase for the three boundary state, the horizon of the BTZ part of the geometry corresponds to the generator $g_1$ (taking the time-reversal to be reflection in the $y$-axis). For the torus wormhole, the generators $g_1,g_2$ correspond to the lengths of the two cycles of the torus behind the horizon, and the horizon itself corresponds to $g_1\cdot g_2 \cdot g_1^{-1} \cdot g_2^{-1}$. Swapping $\alpha\leftrightarrow\tilde{\alpha}$ then gives an equivalent Schottky group, related by an $SL(2,\CC)$ transformation which swaps the two generators. In particular, when $\alpha=\hat{\alpha}$, this becomes an additional automorphism of the Schottky group, corresponding to the torus behind the horizon becoming `square', having equal cycle lengths.

In this case, the boundary terms require more work, since we will need to move the integral on the part of $C_2$ outside the unit circle to the inside, onto $\tilde{C}_2$, using the inversion map $\bar{L}$.
\begin{align}
	\int_{C_2}\phi\; \dd\theta_\infty &= \int_{\tilde{C}_2}\phi\; \dd\theta_\infty - \int_{\bar{L}(\tilde{C}_2)}\phi\; \dd\theta_\infty \\
	&= \int_{\tilde{C}_2}\phi\; \dd\theta_\infty - \int_{\tilde{C}_2}\left(\phi+2\log r\right) \left(2\dd\theta - \dd\theta_\infty\right) \\
	&= \int_{\tilde{C}_2}2\phi\; \dd\theta_0 + \int_{\tilde{C}_2}2\log r \;\dd\left(\theta_0-\theta\right)
\end{align}
Going from the first line to the second requires the equivariance of $\phi$ under $r\mapsto1/r$, as well as some geometry to work out how angles change under inversion ($\theta_\infty\mapsto\pi+2\theta-\theta_\infty$). The last line uses $\theta+\theta_0=\theta_\infty$.

Once again, we use \eqref{TZgeometric}, the symmetry reduction \eqref{symBulk}, and the integral \eqref{intTerm} with $D_2=0,R_2=\tan\alpha$ to get
\begin{align}
	I_{TZ}[\phi] =&  2\int_{\tilde{D}}\dd^2w\left[(\nabla\phi)^2 +e^{2\phi}\right] + 4\int_{\partial\tilde{D}}\phi\; \dd\theta + 4\int_{\partial\tilde{D}} \log|w| \;\dd\theta \\
	  &+4\int_{\tilde{C}_2}2\phi\; \dd\theta_0 + 4\int_{\tilde{C}_2}2\log r \;\dd\left(\theta_0-\theta\right) -4\pi\log\tan^2\alpha \nonumber
\end{align}
and we then integrate by parts, and combine terms to find
\begin{align}
	I_{TZ}[\phi] =&  2\int_{\tilde{D}}\dd^2w\left[-\phi\nabla^2\phi +e^{2\phi}\right]+2\int_{C_1}\phi\;\dd\theta + 2\int_C \phi\;\dd\theta+4\int_{\tilde{C}_2}\phi\; \dd\theta_\infty \nonumber \\
	 &- 8\pi\log  \tan\left(\frac{\hat{\alpha}}{2}\right)  - 8\pi\log\tan\alpha + 8\int_{\tilde{C}_2} \log r \;\dd\theta_0 
\end{align}
This last integral can be evaluated in terms of dilogarithms, or alternatively written as
\begin{equation}
	\int_{\tilde{C}_2} \log r \;\dd\theta_0  = \int_0^{2\alpha} \frac{x}{\sin x} \;\dd x
\end{equation}
which we may easily evaluate numerically.

Now to match moduli between this phase and the other, we again find the lengths $\ell$ of the appropriate cycles fixed by the symmetries, by the integral $\ell=\int e^\phi$. The A-cycles here are those fixed by reflection in the $y$-axis, being the two segments of the axis itself, as well as $C_2$. The B-cycles are fixed by inversion, and are the two arcs of the unit circle $C$ joining $C_2$ and $C_2'$, and the circle $C_1$.

\bibliographystyle{JHEP}
\bibliography{higherz}

\end{document}